\documentclass[reprint,preprintnumbers,amsmath,amssymb,aps,nofootinbib,showkeys,superscriptaddress]{revtex4-2}

\usepackage[utf8]{inputenc}  %
\usepackage{graphicx}  %
\graphicspath{{figs_main/}{figs_SI/}}

\usepackage{xcolor}  %
\usepackage{amsmath}  %
\usepackage{amsfonts}  %
\usepackage{latexsym}  %
\usepackage{amssymb,bbm}  %
\usepackage{bm}  %
\usepackage{verbatim}
\usepackage{dcolumn}  %
\usepackage{subfigure}
\usepackage{soul}
\usepackage{array}
\usepackage{multirow}
\usepackage{colortbl}
\usepackage{dsfont}
\usepackage{upgreek}
\usepackage{enumitem}
\usepackage{float}
\usepackage{afterpage}
\usepackage{verbatim}
\usepackage{listings}
\usepackage{mleftright}

\usepackage[colorlinks=true,allcolors=blue]{hyperref}  %
\usepackage{setspace}
\usepackage{siunitx}
\usepackage{newunicodechar}
\newunicodechar{′}{\ensuremath{'} }
\newunicodechar{−}{\ensuremath{-}}

\newcommand{\app}{\approx}

\newcommand{\beq}{\begin{equation}}
\newcommand{\eeq}{\end{equation}}
                  
\newcommand{\benum}{\begin{enumerate}}
\newcommand{\eenum}{\end{enumerate}}
                    
\newcommand{\bit}{\begin{itemize}}
\newcommand{\eit}{\end{itemize}}

\newcommand{\bea}{\begin{eqnarray}}
\newcommand{\eea}{\end{eqnarray}}

\newcommand{\bev}{\begin{verbatim}}
\newcommand{\eev}{\end{verbatim}}

\newcommand{\noi}{\noindent}

\newcommand{\T}[1]{\textbf{#1}}
\newcommand{\I}[1]{\textit{#1}}

\newcommand{\zfl}[1]{\protect\label{fig:#1}}
\newcommand{\zfr}[1]{\figurename\,\ref{fig:#1}}

\newcommand{\ket}[1]{\left\vert{#1}\right\rangle}

\newcommand{\ba}{\left\{ \begin{array}{lr}}
\newcommand{\ea}{\end{array}\right.}

\newcommand{\blist}[1]{
 \begin{list}{#1}%
 \begin{align}
	 arrow
 \end{align}
 $\checkmark\star
  { \setlength{\itemsep}{3pt}
     \setlength{\parsep}{2pt}
     \setlength{\topsep}{3pt}
     \setlength{\partopsep}{0pt}
     \setlength{\leftmargin}{1em}
     \setlength{\labelwidth}{1em}
     \setlength{\labelsep}{0.5em} } }
\newcommand{\elist}{
  \end{list}  }

\DeclareMathSymbol{\vartheta}{\mathalpha}{letters}{"12}
\DeclareMathSymbol{\theta}{\mathalpha}{letters}{"23}
\DeclareMathSymbol{\phi}{\mathalpha}{letters}{"27}
\DeclareMathSymbol{\varphi}{\mathalpha}{letters}{"1E}

\newcommand{\bef}
{
\begin{figure}[htbp]
\centering
}

\newcommand{\eef}{\end{figure}}

\renewcommand{\figurename}{Fig.}

\newcolumntype{a}{$>{\columncolor{Gray}}c}
\newcolumntype{b}{$>{\columncolor{White}}c}

\newcolumntype{L}[1]{$>{\raggedright\let\newline\\\arraybackslash\hspace{0pt}}m{#1}}
\newcolumntype{C}[1]{$>{\centering\let\newline\\\arraybackslash\hspace{0pt}}m{#1}}
\newcolumntype{R}[1]{$>{\raggedleft\let\newline\\\arraybackslash\hspace{0pt}}m{#1}}

\newcolumntype{P}[1]{>{\centering\arraybackslash}p{#1}}
\newcolumntype{M}[1]{>{\centering\arraybackslash}m{#1}}

\makeatletter
\newif\ifsuppTOC
\suppTOCfalse

\let\orig@addcontentsline\addcontentsline
\renewcommand{\addcontentsline}[3]{%
  \ifsuppTOC
    \edef\tempa{#1}\edef\tempb{toc}%
    \ifx\tempa\tempb
      \orig@addcontentsline{stoc}{#2}{#3}%
    \else
      \orig@addcontentsline{#1}{#2}{#3}%
    \fi
  \else
    \orig@addcontentsline{#1}{#2}{#3}%
  \fi
}

\newcommand{\supplementtableofcontents}{%
  \phantomsection
  \pdfbookmark[1]{Contents}{supp-contents}%
  \begingroup
  \parskip=0pt
  \@starttoc{stoc}%
  \endgroup
}
\makeatother

\newcommand{\beginsupplement}{%
  \suppTOCtrue
  \setcounter{table}{0}
  \renewcommand{\tablename}{Table}
  \renewcommand{\thetable}{T\arabic{table}}%
  \renewcommand{\theHtable}{SI.\arabic{table}}%
  \setcounter{figure}{0}
  \renewcommand{\thefigure}{S\arabic{figure}}%
  \renewcommand{\theHfigure}{SI.\arabic{figure}}%
  \setcounter{page}{1}
  \renewcommand{\figurename}{Fig.}
  \renewcommand{\thesection}{S\arabic{section}}
  \renewcommand{\theHsection}{SI.\arabic{section}}%
  \setcounter{section}{0}
}

\newcommand{\nocontentsline}[3]{}
\newcommand{\tocless}[2]{\bgroup\let\addcontentsline=\nocontentsline#1{#2}\egroup}

\definecolor{Gray}{gray}{0.85}
\definecolor{LightCyan}{rgb}{0.88,1,1}
\definecolor{SITableHeader}{HTML}{F0D5CF}
\definecolor{SITableHighlight}{HTML}{FFF7DE}
\definecolor{SITableRule}{HTML}{5C5C5C}

\newcommand{\affA}{Department of Chemistry, University of California, Berkeley, Berkeley, CA 94720, USA.}
\newcommand{\affB}{Department of Chemical and Biomolecular Engineering, University of California, Berkeley,
CA 94720, USA.}

\newcommand{\affE}{Chemical Sciences Division,  Lawrence Berkeley National Laboratory,  Berkeley, CA 94720, USA.}

\newcommand{\affG}{Department of Pure and Applied Science, University of Urbino Carlo Bo, Urbino, I-61029, Italy.} 
\newcommand{\affH}{Molecular Foundry,  Lawrence Berkeley National Laboratory,  Berkeley, CA 94720, USA.}
\newcommand{\affI}{Department of Physics, Guru Nanak Dev University, Amritsar, Punjab 143005, India.}
\newcommand{\affJ}{Department of Chemistry, University of Texas at Austin, Austin, TX 78712, USA.}
\newcommand{\affK}{Molecular Biophysics and Integrated Bioimaging Division, Lawrence Berkeley National Laboratory, Berkeley, CA 94720, USA.}

\begin{document}
\title{Dense pentacene cocrystal demonstrates room-temperature coherent control}
\author{Noella D'Souza}\affiliation{\affA}\affiliation{\affE}
\author{Guangzhao Chen}\affiliation{\affE}\affiliation{\affH}
\author{Madhur Parashar}\affiliation{\affA}
\author{Wern Ng}\affiliation{\affA}
\author{Brandon Wallace}\affiliation{\affA}
\author{Tatiana Lindahl}\affiliation{\affA}
\author{Tanner S. Volek}\affiliation{\affJ}
\author{Lam Lam}\affiliation{\affA}\affiliation{\affK}
\author{Joseph Garrett}\affiliation{\affA}
\author{Emanuel Druga}\affiliation{\affA}
\author{Sean T. Roberts}\affiliation{\affJ}
\author{Jeffrey Reimer}\affiliation{\affB}
\author{Harpreet Singh}\affiliation{\affA}\affiliation{\affI}
\author{Liang Z. Tan}\affiliation{\affH}
\author{Riccardo Montis}\email{riccardo.montis@uniurb.it}\affiliation{\affG}
\author{Ashok Ajoy}\email{ashokaj@berkeley.edu}\affiliation{\affA}\affiliation{\affE}

\begin{abstract}
Maximizing the number of addressable spins within a fixed volume can improve ensemble quantum sensor sensitivity, but dense packing usually increases dipolar interactions and excited-state transport, shortening coherence and suppressing optical readout. Here we report a 2:1 cocrystal of 6,13-dihydropentacene and pentacene (DHP/Pc) containing 33.3~mol\% pentacene ($3.3\times10^{5}~$ppm; $9.51\times10^{20}~\mathrm{cm}^{-3}$), a volumetric spin-site density more than two orders of magnitude above previous benchmarks, NV-diamond and pentacene-doped \emph{p}-terphenyl (PDP). Despite this density, DHP/Pc exhibits microsecond spin coherence, room-temperature optically detected magnetic resonance and coherent control. Time-resolved measurements indicate that the long-lived triplet population is generated predominantly by intersystem crossing and exhibits strong, non-thermal sublevel polarization. Density-functional theory calculations further suggest that the native cocrystal geometry weakens electronic coupling between neighboring pentacenes, while the higher DHP triplet energy creates a barrier to triplet migration. These results identify molecular packing and coformer triplet energetics as complementary design parameters for preserving coherence in high spin-site density systems, demonstrating cocrystallization as a promising route to dense, optically addressable spin materials for room-temperature ensemble quantum sensing.
\end{abstract}

\maketitle
\pagebreak

Quantum sensors convert perturbations of a quantum state into sensitive measurements of magnetic field, temperature, pressure, and other local quantities \cite{Degen17}. For $N$ independent, optically addressable spins, the standard quantum limit gives a spin projection noise-limited sensitivity that scales as $N^{-1/2}$, provided linewidth, optical contrast, coherence times, polarization, and photon collection efficiencies do not degrade \cite{Taylor_2008,Tesiman_2026}. Increasing the spin-site density in a fixed sensing volume can lower the projection noise associated with measurement, improving the theoretical sensitivity of a quantum sensor \cite{Mitchell_2020}. In practice, gains can be difficult to achieve. High spin density strengthens intersensor interactions, broadens resonances, and shortens coherence \cite{Shinei_2022}; dense chromophore packing can also introduce excited-state transport and quenching mechanisms that reduce optical readout. A central materials challenge is therefore to increase spin-site density without reducing spin coherence.

The nitrogen--vacancy (NV$^{-}$) center in diamond typifies this constraint. Its optical spin polarization, readout, and long room-temperature coherence have enabled the most mature solid-state quantum-sensing platform \cite{Neumann2010,Doherty2013}. In ensembles, however, interactions among NV centers and other paramagnetic defects, particularly substitutional nitrogen, produce a steep concentration--coherence trade-off; high-performance samples therefore commonly employ NV$^{-}$ concentrations from a few to tens of ppm \cite{Acosta_2009,Shinei_2022,Barry2020SensitivityOptimizationNVDiamond}. 
Similar arguments might be expected to hold for photoexcited molecular triplets that can be spin-polarized and optically read out under ambient conditions by optically detected magnetic resonance (ODMR). In the prototypical example of pentacene, at high concentrations, singlet fission (SF) \cite{Bayliss_2019,Unger_2022}, triplet migration \cite{dexter_theory_1953,Kohler_2009}, and triplet--triplet annihilation (TTA) \cite{Poletayev2014} become increasingly competitive, potentially broadening optical transitions and depleting or redistributing the addressable triplet population \cite{Lubert-Perquel2018,Unger_2026,Bossanyi_2021,Bu_Ali_2025}. These processes are traditionally suppressed by diluting pentacene through doping into a host crystal, but this comes at the expense of sensor-site density~\cite{Singh24,Mena24}.

Nonetheless, molecular triplets provide an opportunity to explore a larger design space. Molecular identity and crystal packing jointly define both the spin-bearing site and its local environment, potentially allowing sensor concentration, orientation, and intermolecular interactions to be engineered together \cite{Yu_2021,Han_2026b, Huang_2026}.
Molecular platforms also provide complementary design parameters: chemical substitution can tune spin--orbit and hyperfine interactions \cite{Mann_2025} and correlated triplet-pair dynamics \cite{He_2026,Phansa_2026,Grune_2026}, while crystal packing controls electronic coupling \cite{Lijina_2020,Kolata_2014} and triplet transport \cite{Palmer_2025}. 

We demonstrate here that cocrystallization provides a viable design strategy: sensor molecules densely packed at defined orientations with a coformer of higher triplet-energy can weaken electronic coupling and disfavor triplet transport. We demonstrate that this approach can preserve room-temperature ODMR and coherent control even in a highly packed molecular lattice. Specifically, we investigate a 2:1 cocrystal of 6,13-dihydropentacene and pentacene (DHP/Pc); its lattice contains one pentacene per three molecular sites, yielding 33.3~mol\% ($3.3\times10^{5}~$ppm) pentacene and a site density of $9.51\times10^{20}~\mathrm{cm}^{-3}$ (\zfr{mfig1}B). This exceeds by more than two orders of magnitude the largest volumetric spin-site densities for benchmark cases of NV centers and pentacene-doped \I{para}-terphenyl (PDP) hosts (\zfr{mfig1}E; Table~\ref{tab:funtable}). Despite this dense packing, DHP/Pc retains resolved pentacene optical transitions (\zfr{mfig1}D), spin-polarized triplets (\zfr{mfig2}), and room-temperature ODMR, Rabi oscillations, and echo-based coherent control (\zfr{mfig3}). 

A combination of spectroscopic techniques (transient EPR, FTIR, transient absorption) reveals that intersystem crossing remains the dominant triplet-generation pathway in the cocrystal, while density-functional theory (DFT) calculations suggest that native packing suppresses intermolecular electronic coupling and the DHP--pentacene triplet-energy offset disfavors triplet migration (\zfr{mfig4}). 

\begin{figure*}
        \centering
        \includegraphics[width=1\linewidth]{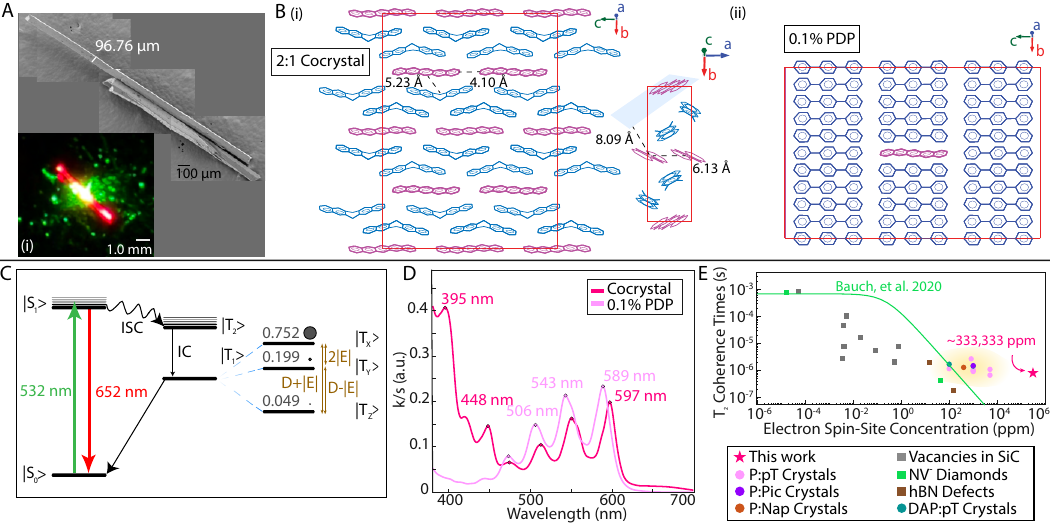}
\caption{\T{Dense DHP/Pc cocrystal and electronic structure.} (A) \I{Morphology and fluorescence.} SEM of DHP/Pc needle (width indicated); (i) optical image of second needle with green laser illumination highlights red fluorescence. (B) \I{Crystal structure.} (i) 2:1 DHP/Pc crystal structure from single-crystal X-ray diffraction. Dashed lines mark nearest pentacene separations along $c$ (4.10~\AA, between nearest two carbon atom-centroid on neighboring pentacenes) and $a$ (6.13~\AA, between molecular centroids); 8.09~\AA{} separation is shortest centroid-to-plane distance to vertical pentacene neighbour. Lattice vectors and molecular axes shown. (ii) Schematic comparison with 0.1\% (w/w) pentacene-doped \emph{p}-terphenyl (PDP); dopant positions schematic, drawing not to scale. (C) \I{Jablonski diagram and triplet sublevel populations.} 532~nm excitation populates vibronically excited $\ket{S_1}$ levels, followed by prompt 652~nm fluorescence or intersystem crossing through $\ket{T_2}$ with preferential $T_{\mathrm{X}}$ population; internal conversion yields $\ket{T_1}$. Inset: $T_{\mathrm{X}}:T_{\mathrm{Y}}:T_{\mathrm{Z}}=0.752:0.199:0.049$ and $D=1376.8~\mathrm{MHz}$, $E=-47.4~\mathrm{MHz}$, extracted from EasySpin fits to EPR data in \zfr{mfig2} (SI Sec.~\ref{SI_EPR}; Table~\ref{tab:EasySpinFit}). (D) \I{Optical absorption.} Solid-state UV/Vis spectra of DHP/Pc and 0.1\% (w/w) PDP. Pentacene singlet transitions at 450--625~nm remain well-resolved and red-shifted without pronounced broadening. (E) \I{Room-temperature, optically detected ensemble $T_2$ times.} Green curve: NV-ensemble model of Ref.~\cite{Bauch2020}, predicting shorter $T_2$ with increasing substitutional-nitrogen spin-bath concentration; continuation beyond reported 0.01--300~ppm range is extrapolated. Symbols: echo-measured $T_2$ versus electron spin-site concentration for NV defects \cite{Bar-Gill_2013,Kucsko_2018,Wang_2020,Jiang_2023}, Si vacancies \cite{Simin_2017,Lekavicius_2022,Kasper_2020,Nagy_2019,wang23,Stuermer_2026,Falk_2013}, hBN defects \cite{Gong_2024,Gottscholl_RT_2021}, DAP-doped crystals \cite{Mann_2025}, pentacene-doped \emph{p}-terphenyl crystals \cite{Singh24,Mena24,Ishiwata2025}, pentacene-doped picene crystals \cite{Huang_2026}, pentacene-doped naphthalene crystals \cite{Huang_2026}, and this work. Unspecified concentrations from the literature were assumed to be mol\%.}
    \zfl{mfig1}
\end{figure*}

\begin{figure*}[t]
  \centering{\includegraphics[width=0.97\textwidth]{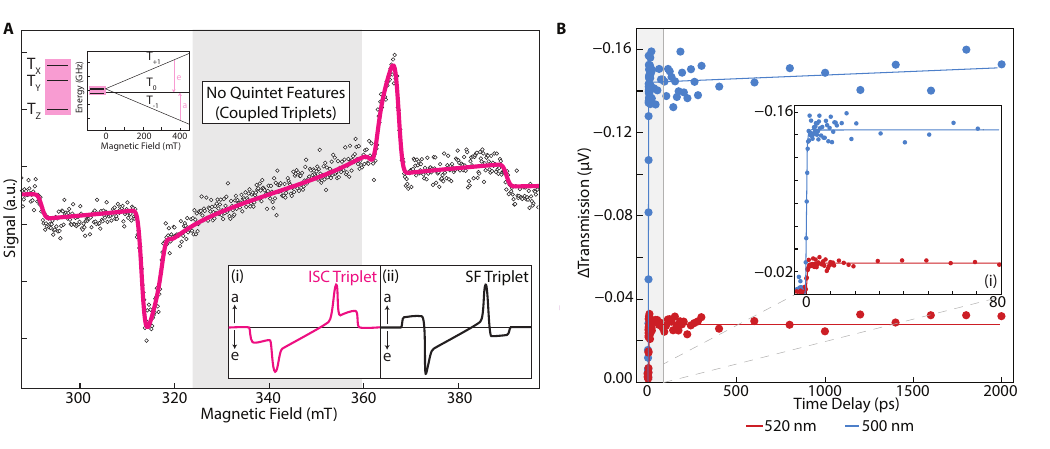}}
\caption{\T{Triplet formation pathway from transient EPR and transient absorption microscopy.} (A) \I{X-band transient EPR.} Powdered DHP/Pc after 100~ns, 527~nm laser pulse; field stepped from 287 to 397~mT; laser pulse energy of 2.5~mJ per pulse and microwave power of $\app 0.01$~mW were used. Emissive (\I{e}) and absorptive (\I{a}) features at $\app 315$ and $\app 365$~mT correspond to $\ket{T_0}\!\rightarrow\!\ket{T_{-1}}$ and $\ket{T_0}\!\rightarrow\!\ket{T_{+1}}$ transitions (inset). \I{eeeaaa} polarization profile supports an ISC-generated long-lived triplet. Shaded region contains no resolved shoulders or peaks assignable to quintet triplet-pair states \cite{Lubert-Perquel2018}. (B) \I{Transient Absorption Microscopy (TAM).} Excited state absorption (ESA) signals from a DHP/Pc cocrystal measured at 500 nm and 520 nm that arise upon photoexcitation of the crystal at 600 nm. The inset (i) shows the ESA signals from 0 – 80 ps. The lack of observable dynamics in the measured ESA signals is consistent with the slow generation of triplet excitons via ISC rather than rapid triplet production via SF.}
  \zfl{mfig2}
\end{figure*}

\vspace{0.5em}
\noi{\normalsize\bfseries Cocrystal Structure and Pentacene Interactions \par}
\noi
The DHP/Pc cocrystal forms as pink and yellow needles during physical vapor transport of pentacene under an inert Ar atmosphere, following thermally-induced disproportionation that generates DHP \cite{Mattheus2002,Roberson2005,Kim2021} (SI Sec.~\ref{SI_CrystalGrowth}; Fig.~\ref{SublSchematic}). Single-crystal X-ray diffraction confirms a monoclinic 2:1 DHP/Pc lattice in which DHP and pentacene occupy defined crystallographic sites and orientations (\zfr{mfig1}B(i)). Scanning electron microscopy (SEM) shows the needle morphology (\zfr{mfig1}A). The representative needle shown in \zfr{mfig1}A is $\sim96~\mu\mathrm{m}$ wide, set predominantly by the duration of sublimation, whereas average needle length is set by the inner diameter of the sublimation tube. SEM experimental parameters and image-analysis details are provided in SI Sec.~\ref{SI_SEM}. The pink and yellow crystals have essentially the same 2:1 lattice, but the yellow form shows weaker pentacene absorption and ODMR, consistent with a lower pentacene fraction; we therefore focus on the pink cocrystal sample for the remainder of these measurements (SI Sec.~\ref{SI_CocrystalSelection}; Figs.~\ref{fig:SI_XRD_Comp}--\ref{fig:SI_ODMR_PinkYellow}; Table~\ref{Tab:XRD_Analysis}). Optical images with green laser illumination (\zfr{mfig1}A(i)) highlight the characteristic red fluorescence of the pink needle (\zfr{Fl}). $^1$H NMR confirms DHP and shows no detectable pentacenequinone in the pink needles (SI Sec.~\ref{SI_NMR}; \zfr{yaynmr}).

Excitation at 532~nm populates vibronically excited levels of the pentacene $\ket{S_1}$ state. Among competing relaxation pathways, $\ket{S_1}$ can emit prompt fluorescence centered at 652~nm or undergo intersystem crossing (ISC) through a nearby triplet state (\zfr{mfig1}C, SI Sec.~\ref{SI_Spectrofluorometry}, \zfr{Fl}). Rapid internal conversion within the triplet manifold then populates the lowest triplet state, $\ket{T_1}$. Zero-field splitting lifts the degeneracy of its three spin sublevels. In contrast to a lattice hosting dilute pentacene, key questions arise regarding the fate of triplet excitons in a cocrystal packed at high spin-site density. High pentacene site density can facilitate triplet-energy migration between chromophores, thereby increasing the probability of encounters between photoexcited triplets. At sufficiently high triplet populations, these encounters can deplete the addressable $\ket{T_1}$ population through TTA~\cite{Poletayev2014}. SF is a distinct pathway in which an excited singlet initially generates a spin-correlated triplet pair, ${}^{1}(\mathrm{TT})$, which can subsequently separate into two $\ket{T_1}$ excitons. The approximate condition $E(S_1)\gtrsim 2E(T_1)$ makes SF energetically favorable; its rate and yield also depend on intermolecular geometry and electronic coupling \cite{Zeng_2014}.

\vspace{0.5em}
\noi{\normalsize\bfseries Results and discussion \par}
\noi

We first compare the XRD-derived pentacene separation in the DHP/Pc cocrystal with a density-derived characteristic spacing in representative 0.1\% (w/w) PDP to anticipate inter-pentacene interaction strength in each system. The smallest distance between pentacene molecules in the cocrystal is 4.10~\AA, measured as the distance between the two carbon atom-centroid on proximal aromatic rings (\zfr{mfig1}B(i)). In comparison, the pentacene number density in the benchmark 0.1\% (w/w) PDP sample corresponds to a characteristic spacing $n^{-1/3}\approx 72$~\AA{} (\zfr{mfig1}B(ii); SI Sec.~\ref{SI_DistanceEstimation}), indicating much closer pentacene packing in the cocrystal.

Comparison of the DHP/Pc cocrystal powder UV/Vis spectrum with that of 0.1\% (w/w) PDP in \zfr{mfig1}D shows no pronounced intermolecular coupling-induced broadening in the cocrystal; pentacene transitions in the 500-625~nm range remain resolved, though slightly red-shifted. Experimental details for the fluorescence, solid-state absorption, and complementary solution-state optical measurements are provided in SI Secs.~\ref{SI_Spectrofluorometry}--\ref{SI_Soln_UVVis}. Lubert-Perquel et al.~\cite{Lubert-Perquel2018} reported that a lower-energy Davydov component emerges in the lowest energy $\ket{S_0}\rightarrow\ket{S_1}$ UV/Vis transition above approximately 1\% pentacene and becomes more pronounced with increasing pentacene concentration in PDP films prepared via thermal vapor deposition. We do not expect to see, nor do we observe, significant Davydov splitting of this transition in the DHP/Pc cocrystal because the parallel transition dipole moment orientation of nearest pentacene molecules means one of the Davydov-split transitions, discussed in \zfr{mfig4}C--D, is optically dark \cite{beljonne_charge-transfer_2013,Zang2017}. Thus, the lowest electronic transition manifests experimentally as a single absorption peak.

\begin{figure*}[t]
  \centering{\includegraphics[width=0.97\textwidth]{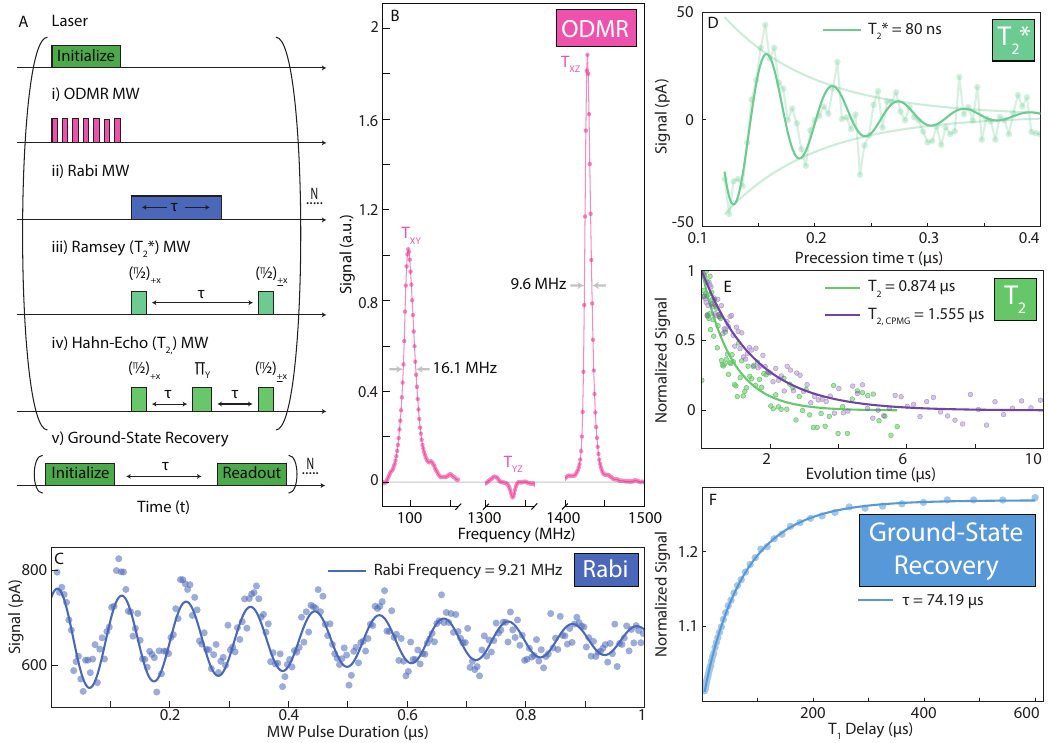}}
\caption{\T{Room-temperature ODMR and coherent control of dense pentacene cocrystal.} (A) Pulse sequences: (i) ODMR, (ii) Rabi, (iii) Ramsey, (iv) Hahn echo, and (v) ground-state recovery. (B) Continuous-wave DHP/Pc ODMR spectrum; resonances at 97, 1334, and 1428~MHz correspond to $T_{\mathrm{XY}}$, $T_{\mathrm{YZ}}$, and $T_{\mathrm{XZ}}$ transitions in photoexcited $\ket{T_1}$. $\app30$~mW laser power at sample and $\app1$~W microwave power for the dataset shown.  (C) Rabi oscillations at 9.2~MHz are shown with solid line fit. (D) Ramsey measurement gives $T_2^{*}=80$~ns. (E) Hahn-echo $T_2=0.874~\mu\mathrm{s}$ extends to $1.6~\mu\mathrm{s}$ with CPMG-4. (F) $\ket{T_1}\rightarrow\ket{S_0}$ ground-state recovery fit to a stretched exponential with time constant $74.19~\mu\mathrm{s}$ and $\beta=0.93$.}
  \zfl{mfig3}
\end{figure*}

\vspace{0.25em}
\noi{\normalsize\bfseries Triplet generation pathways and spin polarization \par}
\noi
Transient electron paramagnetic resonance (EPR) spectra were obtained to probe the mechanism of triplet formation (SF or ISC) and the pentacene triplet sublevel populations in this crystal system. For the remainder of this discussion, we define polarization as the difference between stated triplet sublevel populations. DHP/Pc EPR spectra, shown in \zfr{mfig2}A, were obtained and fit using the \I{pepper} least-squares minimization function for powder spectra in the EasySpin software package \cite{Stoll2006} (SI Sec.~\ref{SI_EPR}). 

Prior work by Tait et al.~\cite{Tait2023} shows that, for the relevant triplet-population mechanisms, ISC-generated triplets commonly produce an emissive/emissive/emissive/absorptive/absorptive/absorptive (eeeaaa) pattern (\zfr{mfig2}A(i)), whereas separated triplets generated through SF can produce an absorptive/emissive/emissive/absorptive/absorptive/emissive (aeeaae) pattern (\zfr{mfig2}A(ii)). The cocrystal data in \zfr{mfig2}A display an eeeaaa pattern consistent with an ISC-generated triplet population. No additional shoulders or peaks assignable to quintet triplet-pair states are resolved in the EPR spectrum \cite{Lubert-Perquel2018,Zang2017}. Quintet formation depends on triplet-pair lifetime, exchange coupling, and spin mixing, so their absence does not by itself exclude SF. The detection of quintet states via EPR is dependent on the instrument timing resolution; our 500~ns time resolution could allow for a minor, short-lived SF contribution to pass unobserved in the EPR window.

We use transient absorption microscopy (TAM) and time-correlated single-photon counting (TCSPC) measurements to access time resolutions sufficient for observation of SF and excited-state decay dynamics in the cocrystal (\zfr{mfig2}B, SI Sec.~\ref{sec:SI_TAM},~\ref{sec:SI_TCSPC}; Fig.~\ref{SI_TAM},~\ref{SI_TCSPC}). TAM transients were measured at probe wavelengths of 500 nm and 520 nm, which were chosen due to their sensitivity to pentacene singlet and triplet excitons, respectively \cite{lee_two_2019,bender_surface_2018,pensack_solution-processable_2017}. If SF occurs, we would expect to see a rapid loss of the TAM signal at 500 nm from singlet excitons and corresponding growth of signal at 520 nm from triplet excitons. In contrast, at both probe wavelengths we observe an instrument-limited rise of an induced absorption signal that remains constant over our 2 ns probe window. The absence of dynamics over our measurement window is consistent with the generation of triplet excitons via ISC, which is expected to occur on much slower timescales than SF. The lack of SF within the cocrystal likely stems from the large spatial separation between pentacene molecules within it that suppresses orbital overlap interactions needed for SF. Indeed, we only observe SF when employing high excitation fluences that act to partially melt the crystal (Fig.~\ref{SI_TAM}B), which presumably allows some pentacene molecules to come into close enough contact to undergo SF. 

TCSPC-measured photoluminescence (PL) decay kinetics provide complementary support for ISC-dominant triplet generation. The cocrystal and 0.1\% (w/w) PDP have comparable fluorescence lifetimes of $7.6\pm0.6~\mathrm{ns}$ and $7.5\pm0.1~\mathrm{ns}$, respectively (Fig.~\ref{SI_TCSPC}). The similarity in fluorescence lifetimes is consistent with the absence of a large, additional singlet-quenching decay channel for the emissive cocrystal population, which is in agreement with prior conclusions about limited SF in this material. The combined time-resolved measurements indicate that the triplets detected in the DHP/Pc cocrystal are primarily generated via ISC.

Fitting the trEPR spectrum yields normalized $T_{\mathrm{X}}$, $T_{\mathrm{Y}}$, and $T_{\mathrm{Z}}$ sublevel populations of $P_{\mathrm{X}}:P_{\mathrm{Y}}:P_{\mathrm{Z}}=0.752:0.199:0.049$ (SI Sec.~\ref{SI_EPR}; Table~\ref{tab:EasySpinFit}). The fitted $T_{\mathrm{Z}}$ population is lower than that reported for pentacene in \emph{p}-terphenyl, where $P_{\mathrm{X}}:P_{\mathrm{Y}}:P_{\mathrm{Z}}=0.76:0.16:0.08$ \cite{Sloop1981}. Consequently, the normalized $X\!Z$ ($Y\!Z$) polarization increases from 0.68 (0.08) in PDP to 0.70 (0.15) in DHP/Pc, indicating an increase in $X\!Z$ ($Y\!Z$) polarization.

Triplet spin polarization is determined by the interplay between sublevel-specific ISC rates and the spin-lattice relaxation between sublevels. Prior calculations show that strong, aromatic =C--H and C--C out-of-plane (oop) vibrations in pentacene mix $\pi$ and $\sigma$ orbitals to enhance spin-vibronic coupling to the $T_{2,\mathrm{X}}$ and $T_{2,\mathrm{Y}}$ sublevels, while contributing little to $T_{2,\mathrm{Z}}$. This preference can result in lower relative $P_{\mathrm{Z}}$ population if retained during internal conversion \cite{Sakamoto2023,penfold_spin-vibronic_2018}. Ref.~\cite{Sakamoto2023} identifies pentacene out-of-plane modes near 761 and $807~\mathrm{cm}^{-1}$ with strong spin-vibronic coupling to the $T_X$ state. We find these modes enhanced in our normalized solid-state IR spectra in the aromatic =C--H oop band at $837~\mathrm{cm}^{-1}$ (SI Sec.~\ref{SI_SS_FTIR}; Fig.~\ref{SS_FTIR}A), yet the $T_X$ polarization is reduced compared to the PDP system. These contrasting results highlight how modifications to both intrinsic vibrational modes and spin-lattice relaxation rates need to be considered in optimizing cocrystals for quantum sensing applications.

\begin{figure*}[t]
  \centering{\includegraphics[width=0.97\textwidth]{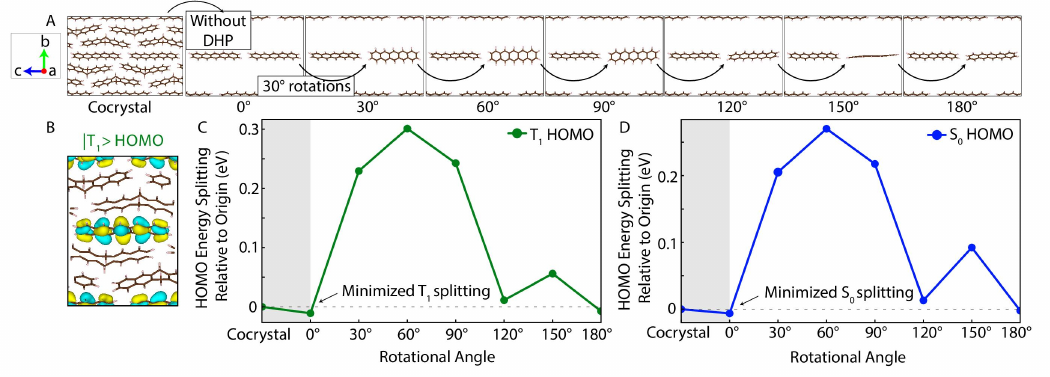}}
  \caption{\T{DFT links suppressed electronic coupling at high pentacene density to pentacene dimer configuration.} (A) \I{Unit-cell geometries} used to rotate horizontal nearest-neighbour pentacene dimer. (B) \I{Representative $T_1$ HOMO distribution} for pentacene in cocrystal geometry. (C,D) \I{Calculated $T_1$ and $S_0$ HOMO energy splitting versus rotation angle.} Native cocrystal angle lies near global splitting minimum in both states, consistent with weak intermolecular electronic coupling and retained ODMR in dense pentacene ensemble.}
  \zfl{mfig4}
\end{figure*}

\vspace{0.25em}
\noi{\normalsize\bfseries Room-temperature ODMR and spin coherence \par}
\noi

Ensemble ODMR spectra of the pentacene triplet transitions in the DHP/Pc cocrystal are shown in \zfr{mfig3}B (SI Sec.~\ref{SI_ODMR}), with the ODMR pulse sequence in \zfr{mfig3}A(i). $T_{\mathrm{XY}}$, $T_{\mathrm{YZ}}$, and $T_{\mathrm{XZ}}$ resonances are at 97, 1334, and 1428~MHz, respectively. The cocrystal's trEPR-derived zero-field-splitting parameters, $D=1376.8~\mathrm{MHz}$ and $E=-47.4~\mathrm{MHz}$, are modified from the values reported for a PDP host, $D=1395.27~\mathrm{MHz}$ and $|E|=53.51~\mathrm{MHz}$ \cite{Yang00}. The power-unbroadened XY linewidth at the stated optical power is $10.2\pm0.6$~MHz (SI Sec.~\ref{SI_Linewidth}; \zfr{SI_MWDependence}), approximately $5\times$ the 1.9~MHz power-unbroadened XY linewidth reported for PDP in Ref.~\cite{Singh24}, likely arising from higher baseline dipolar broadening from the increased proximity of pentacene molecules in this system. The power-unbroadened XZ linewidth at the stated optical power is narrower, at $7.55\pm0.20$~MHz, and was therefore used for the DC-sensitivity estimate. Coherent-control measurements were performed on the YZ transition (SI Secs.~\ref{SI_Coherence}--\ref{SI_Sensitivity}). As expected at high concentrations, the lineshape is approximately Gaussian \cite{Kittel1953}, most clearly visible in the $T_{\mathrm{XZ}}$ peak.

The DC field sensitivity was estimated following the method employed in Ref.~\cite{Singh24}, using $\eta_{\mathrm{DC}}=\sigma\sqrt{\tau}/[\frac{\mathrm{d}S}{\mathrm{d}F}\gamma_{\mathrm{e}}]$, where $\sigma$ is the standard deviation of the off-resonance signal, $\tau$ is the integration time, $\mathrm{d}S/\mathrm{d}F$ is the maximum slope in the ODMR signal, and $\gamma_{\mathrm{e}}$ is the electron gyromagnetic ratio. We estimate a sensitivity of $157~\mathrm{nT}/\sqrt{\mathrm{Hz}}$ from the $T_{XZ}$ transition ODMR slope at zero applied field (SI Sec.~\ref{SI_Sensitivity}), for an effective cubic excitation volume with linear dimension $\approx116~\mu\mathrm{m}$. This value is not optimized and significant room for improvement remains. For example, optical saturation was not reached here because higher CW powers risked heating and damaging the cocrystal. Co-optimized, high-power pulsed optical and microwave driving of the dense pentacene ensemble, together with rapid heat dissipation via a heat-conductive substrate such as sapphire, and with encapsulation, will enable improved ODMR features and enhanced sensitivity in this cocrystal.

\zfr{mfig3}C shows optically accessible coherent control of the electron spin ensemble via Rabi oscillations at 9.2~MHz with the pulse sequence shown in \zfr{mfig3}A(ii). An optically detected transverse electron-spin coherence time $T_2^{*}=80~\mathrm{ns}$ is extracted from detuned Ramsey fringes in \zfr{mfig3}D with the pulse sequence in \zfr{mfig3}A(iii). The shortened $T_2^{*}$ relative to the PDP sample is consistent with increased inhomogeneous dephasing. This may arise from hyperfine and electron--electron dipolar coupling at higher densities, environmental inhomogeneity arising from two inequivalent pentacene lattice sites, and, to a lesser extent, increased strain in the smaller cocrystal lattice, compared with pentacene probed in the bulk PDP crystal of Ref.~\cite{Singh24}. We measure a $\ket{T_1}\rightarrow\ket{S_0}$ ground-state recovery time of $74.19~\mu\mathrm{s}$, with stretched exponent $\beta=0.93$ (\zfr{mfig3}F) using the sequence shown in \zfr{mfig3}A(v).  This value is enhanced, compared with PDP~\cite{Singh24}. We hypothesize that this arises from a reduction in low-frequency molecular librations upon replacing the \emph{p}-terphenyl host with the cocrystallizing DHP molecule \cite{miyokawa_electron_2026,Kohda1982}. 

Echo-measured $T_2$ extends the spin coherence time to $0.874~\mu\mathrm{s}$ by refocusing inhomogeneity-induced dephasing using the Hahn-echo sequence in \zfr{mfig3}A(iv). An approximate twofold coherence improvement to $1.6~\mu\mathrm{s}$ is observed with dynamical decoupling via the CPMG-4 sequence (\zfr{mfig3}E). Experimental details are provided in SI Sec.~\ref{SI_Coherence}. Notably, these $T_2$ times are of the same order of magnitude as $T_2=2.7~\mu\mathrm{s}$ and $T_2^{\mathrm{DD}}=18.4~\mu\mathrm{s}$ reported for PDP \cite{Singh24}. \zfr{mfig1}E summarizes the central concentration--coherence result: DHP/Pc retains microsecond-scale ensemble coherence at 333{,}333~ppm molecular-site concentration. Table~\ref{tab:funtable} shows a volumetric spin-site density more than two orders of magnitude above the NV-diamond and PDP benchmarks compiled here. For comparison, the solid curve shows the prediction for NV-ensembles following Ref.~\cite{Bauch2020}, where $T_2$ decreases with increasing substitutional-nitrogen spin-bath concentration. Recognizing that molecular-site concentration is not the instantaneous triplet-spin density, DHP/Pc lies above the extrapolated diamond-specific trend. 

\vspace{0.5em}
\noi{\normalsize\bfseries Effect of crystal structure on electronic coupling and triplet migration \par}
\noi

Pure pentacene crystals are known to have efficient triplet migration and annihilation \cite{Poletayev2014,Bossanyi_2021,Unger_2022}, quenching the triplet state and complicating their use as ODMR-based quantum-sensing materials. Why is the dense 2:1 cocrystal system not subject to the same fate? We use DFT calculations to address this question in two ways: evaluating energetic barriers to triplet migration and assessing the strength of electronic coupling between nearest-neighbor pentacene dimers in their cocrystal configuration (SI Sec.~\ref{SI_DFT}). 

Triplet migration between adjacent molecules is generally facilitated by $\pi$-orbital overlap and energetically close triplet levels \cite{Kohler_2009}. The cocrystal structure positions pentacene to avoid overlapping $\pi$ orbitals with both nearby pentacene molecules and DHP, as they are $>8$~\AA{} and 5.2~\AA{} apart, respectively, in the $\pi$-stacking direction (see \zfr{mfig1}B(i)). The calculations also place the DHP-localized triplet ($2.98~\mathrm{eV}$) far above the pentacene-localized triplet ($0.78~\mathrm{eV}$), yielding an uphill triplet-energy difference that strongly disfavors pentacene to DHP triplet transfer and corresponding triplet migration via DHP sites (Fig.~\ref{DFT_DHPPT}). A similar approach leveraging energetic barriers to block triplet formation in organic cocrystals \cite{Ye_2018} and triplet migration in pentacene-based organic photovoltaics \cite{Tabachnyk_2013} has been employed to optimize singlet-exciton formation and collection efficiencies, respectively. The estimated energy barrier between pentacene and DHP reported here ($2.2~\mathrm{eV}$) is higher than the energy barriers reported for successful triplet blocking in prior work.

We assess the strength of intermolecular electronic coupling by calculating the orbital energy splitting of the pentacene singlet and triplet HOMO states to model Davydov splitting. The triplet HOMO distribution used in this analysis is shown in \zfr{mfig4}B. The orbital splitting conformational dependence is probed qualitatively by rotating one molecule of each pentacene dimer through fixed angular increments relative to its crystallographic orientation and calculating the HOMO energy difference between two pentacene molecules in the supercell for horizontal (\zfr{mfig4}A; Fig.~\ref{DFT_Horizontal}) and vertical dimer configurations (Fig.~\ref{DFT_Vertical}). Nearest-plane separations for the two horizontal dimers are 4.10 and 6.13~\AA{}, while the vertical centroid-to-plane separation is 8.09~\AA{} (see \zfr{mfig1}B(i)). Removing DHP molecules to reduce computational cost had little effect on the calculated splitting at the native geometry (compare Cocrystal and $0^\circ$ points in \zfr{mfig4}C--D).

Importantly, we find that the native orientation between horizontal and vertical pentacene dimers lies near a global energy minimum in the orbital energy splitting curve in both the $T_1$ and $S_0$ manifolds over the sampled rotation angles (\zfr{mfig4}C--D; Figs.~\ref{DFT_Vertical}--\ref{DFT_Horizontal}, panels B--C). These trends are consistent with minimal intermolecular coupling in the lowest triplet and singlet electronic states, which may favor triplet localization and help preserve observable spin coherence \cite{Palmer_2025,Han_2026b}. Comparing pentacene's orbital spread along all molecular axes supports this finding as well. Greater orbital localization along the short in-plane and out-of-plane molecular axes for pentacene in the cocrystal, compared with pure pentacene, is consistent with comparatively reduced intermolecular electronic interactions along these axes (Fig.~\ref{DFT_Spread}).

A distinct Davydov component is not resolved in the experimental UV/Vis spectrum (\zfr{mfig1}D), since one exciton combination of the nearly parallel horizontal dimer is optically weak for the $\ket{S_0}\rightarrow\ket{S_1}$ transition. The calculated orientational dependence nevertheless qualitatively resembles reported coupling trends for crystalline pentacene \cite{Zeng_2014}.

\vspace{0.5em}
\noi{\normalsize\bfseries Discussion and Outlook \par}
\noi
Our results show that molecular cocrystals provide a largely unexplored route to dense, optically addressable electron spin ensembles at room temperature~\cite{attwood_probing_2024,Palmer_2024,Palmer_2025}; the lattice fixes spin-site concentration, while relative molecular orientation can suppress intermolecular electronic coupling that would otherwise extinguish triplet-spin coherence at stoichiometric densities. The DHP/Pc cocrystal exemplifies this by demonstrating room temperature ODMR and microsecond-long spin coherence times at remarkably high (33.3 mol\%) pentacene spin-site concentrations (see \zfr{mfig1}E).

The DHP/Pc material inspires multiple approaches to general fabrication of dense optically-addressable, room temperature-coherent, solid-state spin ensembles. First, cocrystallization offers a modular strategy for varying (i) energetic barriers to triplet migration and sensor packing, (ii) triplet state localization and spin coherence properties \cite{Palmer_2024}, (iii) vibronic structure to tune intersystem crossing rates and sublevel selectivity \cite{Sakamoto2023} to modes which may be otherwise challenging to modify chemically, and (iv) the zero-field splitting tensor, which sets ODMR transition frequencies \cite{Yang00}. A practical next step is to build and benchmark a small library of pentacene-based cocrystals using rapid sublimation-based crystal growth techniques and varying coformers to map modifications to triplet quantum yield, electron spin polarization, spin coherence, and density-dependent dephasing. DFT screening can assist in prioritizing cocrystal syntheses with minimal electronic coupling and favorable triplet energetics.

Second, the needle-like single-crystal morphology in \zfr{mfig1}A suggests straightforward integration with optical and microwave hardware. Oriented crystals could be embedded in resonators or coupled to waveguides to enhance pump absorption and fluorescence collection, as in Ref. \cite{Faraon_2012}, directly translating to gains in magnetometry or thermometry ~\cite{Singh24,Mena24,Li_2026,Singh2024_PT}. Systematic studies of cocrystal coherent control as a function of excitation fluence, temperature, and applied field will be important for identifying the onset of density-dependent broadening and for separating dipolar dephasing from triplet-transport-related signal loss \cite{Palmer_2024}.

Finally, dense triplet ensembles are attractive beyond sensing. The same non-equilibrium triplet polarization that underpins ODMR-based quantum sensors can be harnessed for triplet dynamic nuclear polarization \cite{Tateishi2014,Quan2019}. Cocrystal engineering may allow co-optimization of optical absorption, triplet yield, and nuclear-spin coupling for such applications. More broadly, the ability to tune sensor packing and energetics to maintain high spin-site density provides a pathway to organic quantum materials whose function can be modulated by both supramolecular structure and molecular identity.

\section*{Methods}\label{sec:Methods}
\vspace{0.5em}
\noi{\normalsize\bfseries Cocrystal growth and structural characterization\par}
\noi
Pink 2:1 DHP/Pc cocrystals were grown by physical vapor transport of pentacene (Sigma-Aldrich, 99\%) in a home-built horizontal gradient furnace. Approximately 120~mg of pentacene was heated to 330~$^\circ$C under 99.999\%-purity Ar flowing at 0.4~mL\,s$^{-1}$; DHP/Pc needles deposited near 170~$^\circ$C after in-situ vapor-phase disproportionation of pentacene. Pink and yellow products were compared by single-crystal X-ray diffraction, with data from four pink and three yellow needles analysed using Mercury. Crystal morphology was examined without conductive coating in a Zeiss Crossbeam 550 scanning electron microscope at 0.7~kV and 100~pA, and needle dimensions were extracted in Fiji. Solution-state $^1$H NMR spectra were acquired at 400~MHz in CDCl$_3$ to verify DHP and test for pentacenequinone. Representative morphology and unit-cell structure are shown in \zfr{mfig1}A--B. Growth, SEM analysis and NMR are detailed in SI Secs.~\ref{SI_CrystalGrowth}, \ref{SI_CocrystalSelection}, \ref{SI_SEM}, and~\ref{SI_NMR}.

\vspace{0.5em}
\noi{\normalsize\bfseries Steady-state and time-resolved spectroscopies\par}
\noi
Fluorescence spectra were recorded under 532~nm excitation using a Tecan Spark plate reader. Diffuse-reflectance UV/Vis spectra of powdered samples were acquired using a Shimadzu UV-2600 spectrophotometer and converted to pseudo-absorbance using the Kubelka--Munk transform; solution-state spectra in toluene were acquired using a Varian Cary 50 spectrophotometer. ATR-FTIR spectra were recorded on a Bruker Vertex 70 from 400 to 4500~$\mathrm{cm}^{-1}$ at 4~$\mathrm{cm}^{-1}$ resolution with 100 scans per dataset, averaged over three replicate measurements and baseline-corrected with \I{msbackadj} in MATLAB. Acquisition and analysis procedures for steady-state spectroscopies are provided in SI Secs. ~\ref{SI_Spectrofluorometry}--\ref{SI_Soln_UVVis} and \ref{SI_SS_FTIR}.

Fluorescence lifetimes were measured by time-correlated single-photon counting under 520~nm excitation. 650~nm long-pass emission was detected using a superconducting nanowire single-photon detector; linear fits to log-intensity decays were used to extract the lifetimes. 
Transient absorption measurements were performed using 600 nm pump pulses to drive a Ti:sapphire regenerative amplifier (Coherent, RegA 9000) coupled to an optical parametric amplifier (OPA). A fraction of the OPA’s 600 nm was focused into a sapphire plate to generate a supercontinuum that was spectrally filtered to yield probe pulses centered at either 500 nm or 520 nm. Data plotted in \zfr{mfig2}B was recorded using a pump fluence of $62~\mu J/cm^2$ and pump spot diameter of $\app1~\mu$m. Detailed TCSPC and TAM acquisition and analysis procedures are provided in SI Secs. ~\ref{sec:SI_TCSPC} and ~\ref{sec:SI_TAM}.

\vspace{0.5em}
\noi{\normalsize\bfseries Transient electron paramagnetic resonance\par}
\noi
Powdered cocrystal was loaded into a 3~mm borosilicate tube and measured using a home-built X-band transient-EPR spectrometer with IQ homodyne detection. Samples in a loop-gap resonator were excited using 100~ns, 527~nm laser pulses with a pulse energy of 2.5~mJ while a 9.565~GHz continuous microwave field was applied. The time-domain response was integrated from 1.5 to 3.4~$\mu$s after excitation and recorded as a function of magnetic field; 50 field sweeps were averaged. The powder spectrum was fit with a least-squares minimization using the \textit{pepper} function in EasySpin to extract zero-field-splitting parameters and normalized triplet-sublevel populations; resulting spectrum and sublevel analysis are shown in \zfr{mfig2}A (SI Sec.~\ref{SI_EPR}).

\vspace{0.5em}
\noi{\normalsize\bfseries ODMR, coherent control and sensitivity\par}
\noi
Room-temperature ODMR was measured under continuous-wave 532~nm excitation. Red photoluminescence was collected through a microscope objective, separated from the excitation path and detected using an avalanche photodiode. Microwaves generated by a Tabor Proteus arbitrary-waveform transceiver were amplitude-modulated at 1~kHz, amplified and delivered using a two-turn Helmholtz coil; the modulated photoluminescence was detected with a lock-in amplifier. Individual cocrystal needles were mounted at the end of a capillary and centered in the microwave coil. Frequency-swept spectra and microwave power-dependent linewidths were acquired using this apparatus (SI Sec.~\ref{SI_ODMR}). Rabi, Ramsey, Hahn-echo and CPMG-4 measurements on $T_{\mathrm{YZ}}$ used synchronized optical and phase-controlled microwave pulses delivered through a 50~$\Omega$ loop resonator. Emission was recorded on a photodiode using lock-in or digitized acquisition. Ground-state recovery was measured using the two-pulse optical sequence in \zfr{mfig3}A(v), with sequence timings and acquisition conditions given in SI Sec.~\ref{SI_Coherence}. Pulse sequences and corresponding data are shown in \zfr{mfig3}A--F. DC sensitivity was estimated from the $T_{\mathrm{XZ}}$ resonance using the off-resonance noise, integration time, maximum ODMR slope and electron gyromagnetic ratio as detailed in SI Sec.~\ref{SI_Sensitivity}.

\vspace{0.5em}
\noi{\normalsize\bfseries Density-functional theory calculations\par}
\noi
Spin-polarized DFT calculations were performed in VASP~\cite{Kresse1996} using a plane-wave basis, projector-augmented-wave potentials and the PBE generalized-gradient approximation\cite{Perdew1996}, with $\Gamma$-point Brillouin-zone sampling. The energy convergence thresholds for the electronic SCF cycle and geometry optimization were set to $10^{-6}$ and $10^{-5}$~eV, respectively. Orbital occupations were constrained to represent selected singlet and triplet configurations, and geometries were optimized using a conjugate-gradient algorithm. 
The DHP/Pc unit cell from Ref. \cite{Mattheus2002} contained four DHP molecules and two pentacene molecules, with lattice parameters $a = 6.163$~\AA, $b = 21.801$~\AA, $c = 16.069$~\AA, $\alpha = \gamma = 90^\circ$, and $\beta = 93.730^\circ$.
The simulation cell of 6.25~mol\% pentacene-doped \emph{p}-terphenyl contained one pentacene molecule and 15 \emph{p}-terphenyl molecules, with lattice parameters $a = 16.212$~\AA, $b = 11.226$~\AA, $c = 27.226$~\AA, $\alpha = \gamma = 90^\circ$, and $\beta = 92.020^\circ$. The pure pentacene unit cell contained two pentacene molecules, with lattice parameters $a = 6.239$~\AA, $b = 7.636$~\AA, $c = 14.330$~\AA, $\alpha = 76.978^\circ$, $\beta = 88.136^\circ$, and $\gamma = 84.415^\circ$.
Calculations compared DHP/Pc with pure pentacene and dilute and dense pentacene-doped \emph{p}-terphenyl, and evaluated state energies, triplet-energy differences, orbital localization, and the angular dependence of the pentacene-dimer HOMO-level splitting, used here as a qualitative proxy for intermolecular electronic coupling.
The orbital energy splitting is evaluated from the energy range spanned by the pentacene-derived frontier orbitals. In \zfr{mfig4}, the supercell contains four pentacene molecules; therefore, the HOMO-derived splitting is calculated as $E(\mathrm{HOMO}) - E(\mathrm{HOMO}-3)$.
Similarly, the HOMO-derived splitting is defined as $E(\mathrm{HOMO}) - E(\mathrm{HOMO}-1)$ in \zfr{DFT_Vertical} and $E(\mathrm{HOMO}) - E(\mathrm{HOMO}-3)$ in \zfr{DFT_Horizontal}. Complementary DFT normal-mode calculations were compared with the FTIR spectra in Fig.~\ref{SS_FTIR}A(iii). Further computational details are provided in SI Sec.~\ref{SI_DFT}.

\vspace{0.5em}
\noi{\normalsize\bfseries Acknowledgements \par}
\noi
The authors gratefully acknowledge discussions with A. Sau and E.E. Brown, experimental assistance from L. Jayasinghe, P. Barzova, G. Brown, and M. Lloyd, and funding from NNSA (LB24-NV center 13C quantum sensor-PD3Ta, LB26-nuclear-spin quantum sensing-PD3Td), U.S. DOE (DE-FOA-0001968, DE-SC0025524), ONR (N00014-20-1-2806), AFOSR YIP (FA9550-23-1-0106), and the Sloan and Dreyfus Foundations. The authors gratefully acknowledge Nick Settineri and the UC Berkeley CheXray Facility for 2:1 DHP/Pc XRD data collection and crystal structure determination. Work at the Molecular Foundry was supported by the Office of Science and Office of Basic Energy Sciences of the U.S. Department of Energy under Contract No. DE-AC02-05CH11231. Computational work used resources of the National Energy Research Scientific Computing Center (NERSC), a Department of Energy User Facility, using NERSC allocation m4269 under the award BES-ERCAP0032327. Work at the University of Texas at Austin was supported by the National Science Foundation under award CHE-2544563.

\vspace{0.5em}

\vspace{-5mm} 
\bibliography{Cocrystal_Pentacene_bib}
\clearpage

\beginsupplement
\setcounter{topnumber}{4}
\setcounter{dbltopnumber}{3}
\setcounter{totalnumber}{6}
\renewcommand{\topfraction}{0.95}
\renewcommand{\dbltopfraction}{0.95}
\renewcommand{\textfraction}{0.05}
\renewcommand{\floatpagefraction}{0.80}
\renewcommand{\dblfloatpagefraction}{0.80}
\makeatletter
\setlength{\@fptop}{0pt}
\setlength{\@fpbot}{0pt plus 1fil}
\setlength{\@dblfptop}{0pt}
\setlength{\@dblfpbot}{0pt plus 1fil}
\makeatother

\begin{widetext}
\begin{center}
\textbf{\large\textit{Supplementary Information}}\\[0.5em]
\textbf{Dense pentacene cocrystal demonstrates room-temperature coherent control}\\[1em]
\textbf{CONTENTS}
\end{center}
\end{widetext}

\setcounter{tocdepth}{2}
\supplementtableofcontents

\section*{Guide to the supplementary information}
\label{sec:SI_organization}

This Supplementary Information (SI) is organized in four parts. We first describe material preparation and establish why the pink DHP/Pc cocrystal is used for measurements. SI Sec.~\ref{SI_CrystalGrowth} details physical-vapor-transport crystal growth and compares the pink and yellow products. The two forms have the same 2:1 lattice, but the weaker pentacene absorption and ODMR of the yellow form are consistent with a slightly lower pentacene concentration (Figs.~\ref{fig:SI_XRD_Comp}--\ref{fig:SI_ODMR_PinkYellow}; Table~\ref{Tab:XRD_Analysis}). SI Sec.~\ref{SI_SEM} documents the SEM measurement and image analysis used to quantify the needle dimensions. SI Sec.~\ref{SI_DistanceEstimation} derives a density-based characteristic pentacene spacing of $72.33$~\AA{} in 0.1\% (w/w) PDP. SI Sec.~\ref{SI_Density} benchmarks spin-site density across optically-readout quantum sensing materials shown in main paper Fig. 1 and shows that the DHP/Pc cocrystal reaches $9.514\times10^{20}~\mathrm{cm}^{-3}$ (Table~\ref{tab:funtable}).

We next spectroscopically characterize the sample and determine the triplet-generation pathway. SI Sec.~\ref{SI_Spectrofluorometry} identifies a fluorescence maximum at 652~nm, and SI Sec.~\ref{SI_SS_UVVis} shows that the solid-state pentacene absorptions remain resolved without pronounced broadening. Solution-state UV/Vis provides an independent pentacene-content estimate of approximately 27\% (SI Sec.~\ref{SI_Soln_UVVis}). Transient EPR and EasySpin fitting identify an intersystem crossing (ISC)-dominated polarization pattern and quantify the triplet-sublevel populations (SI Sec.~\ref{SI_EPR}; Table~\ref{tab:EasySpinFit}). Low triplet excited state absorption in ps-timescale transient absorption microscopy is consistent with a primary ISC triplet generation mechanism (SI Sec.~\ref{sec:SI_TAM}). Comparable $\app7.6$~ns fluorescence lifetimes in DHP/Pc and 0.1\% (w/w) PDP indicate that singlet fission is not appreciably enhanced in the cocrystal (SI Sec.~\ref{sec:SI_TCSPC}), while $^1$H NMR confirms DHP and shows no detectable pentacenequinone (SI Sec.~\ref{SI_NMR}).

Subsequently, we discuss ODMR, coherent-control, and sensing methods. SI Sec.~\ref{SI_ODMR} describes the optical and microwave apparatus, sample mounting, and microwave-power dependence of the ODMR linewidths; the $T_{\mathrm{XZ}}$ transition provides the narrowest high signal-to-noise ratio (SNR) resonance for sensing. SI Sec.~\ref{SI_Coherence} details the pulse-control and detection procedures supporting the Rabi, Ramsey, echo, CPMG-4, and ground-state-recovery measurements. SI Sec.~\ref{SI_Sensitivity} then gives the DC-sensitivity calculation and the best measured value of $157~\mathrm{nT}/\sqrt{\mathrm{Hz}}$.

Finally, SI Sec.~\ref{SI_DFT} details constrained-DFT energy-level calculations across the DHP/Pc cocrystal, dilute and dense PDP, and pure pentacene. We estimate triplet-energy, Davydov-splitting, structural-energy, and orbital-spread calculations across samples. These calculations identify a $\app2.2$~eV barrier to DHP-mediated triplet migration and weak intermolecular electronic coupling in native pentacene-dimer geometries. These conclusions favor ISC in the cocrystal but cannot alone exclude singlet fission, motivating the time-resolved measurements in SI Sec.~\ref{sec:SI_TCSPC}.

\section{Crystal Growth}\label{SI_CrystalGrowth}
A sublimation tube is prepared by rinsing a 30~cm-long, 1.2~mm-thick glass tube with an inner diameter of $\app1$~cm sequentially with deionized water and acetone. Pentacene (Sigma Aldrich, 99\%) is transferred into the sublimation tube. The loaded sublimation tube is inserted into a home-built horizontal sublimation furnace (\zfr{subl}A) consisting of copper-wire-based heating coils wound with increasing spacing to generate the temperature gradient described in \zfr{subl}B. The temperature profile in \zfr{subl}B was measured without Ar flowing through the furnace, and is therefore shifted uniformly upward relative to the 330~$^\circ$C hot zone and 170~$^\circ$C deposition zone observed during cocrystal growth via thermal camera. An initial mass of $\app120$~mg stock pentacene is positioned across 2-4 cm in the indicated hot-zone region. The sublimation tube is then connected to a 99.999\%-purity argon-gas line and purged for 1~h to remove O$_2$ from the atmosphere.

\begin{figure*}[t]
    \centering
    \includegraphics[width=0.72\textwidth]{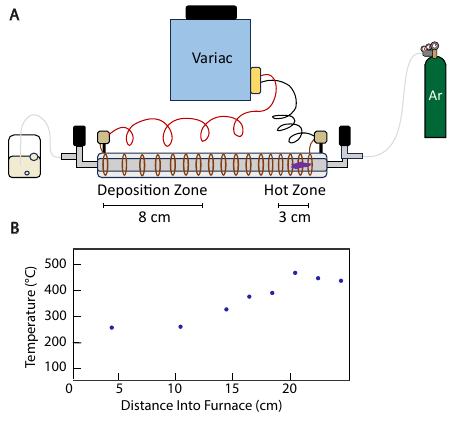}
\caption{\T{Physical vapor transport growth of the DHP/Pc cocrystal.} (a) \I{Horizontal gradient furnace} used for crystal growth. The 3 cm-long hot zone is loaded with pentacene (purple) and heated through Variac-controlled (red/black leads) resistive heating coils. Ultra-high-purity Ar (green; 24~mL\,min$^{-1}$) transports sublimed species from the hot zone through the temperature gradient to the 8 cm-long deposition zone, where pink DHP/Pc needles grow. Exhaust gas is bubbled through mineral oil (beige). (b) \I{Representative temperature profile} along the furnace, obtained from a thermocouple placed in the furnace at varying distances, starting at the right hand side near the hot zone.}
    \label{SublSchematic}
    \zfl{subl}
\end{figure*}

A Variac variable transformer supplies the current required to achieve a hot-zone temperature of 330~$^\circ$C. At this temperature, pentacene sublimes and undergoes a vapor-phase disproportionation reaction to form dihydropentacene (DHP) \cite{Roberson2005}. The Ar flow rate is adjusted to 0.4~mL\,s$^{-1}$ to transport DHP and pentacene vapor through the temperature gradient into the deposition zone. At a distance of 8~cm into the furnace, corresponding to a temperature of 170~$^\circ$C, a 2:1 DHP:pentacene cocrystal deposits as long pink needles. Additional products, listed in order of deposition from the hot zone to the deposition zone, are polycyclic aromatic hydrocarbons, purified pentacene, and a yellow-white 2:1 cocrystal also containing DHP and pentacene.

\subsection{Cocrystal Selection}\label{SI_CocrystalSelection}
\begin{figure*}[t]
  \centering
  {\includegraphics[width=0.88\textwidth]{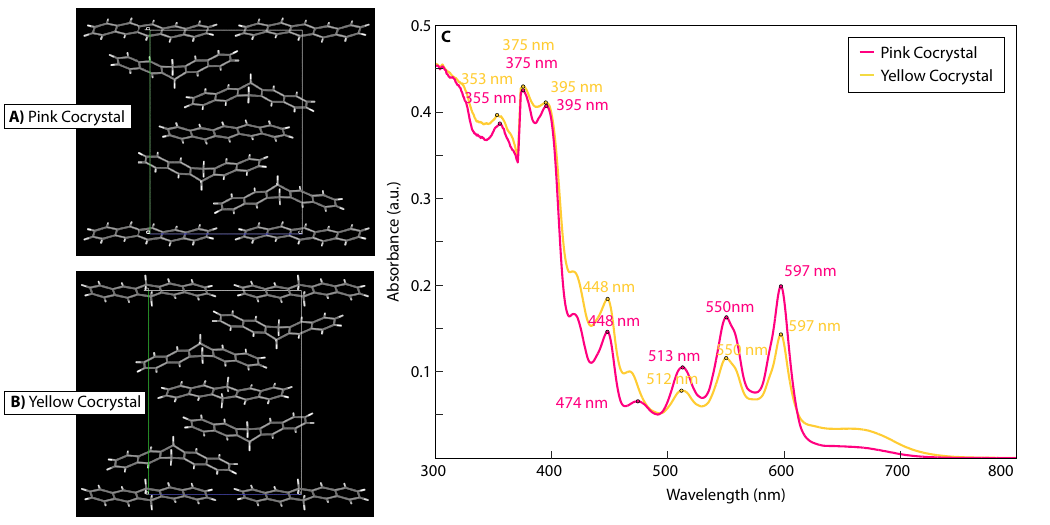}}
  \caption[Cocrystal unit cells]{\T{Unit Cells} of (A) \I{pink}, and (B) \I{yellow 2:1 6,13-dihydropentacene:pentacene cocrystals}. View of unit cell is down the b axis. Crystals grow along the a axis. (C) \T{Diffuse Reflectance Powder UV/Vis} of yellow and pink 2:1 6,13-DHP/Pc cocrystals shows retained pentacene singlet transitions in the 500-625~nm region and different relative intensities of pentacene / non-pentacene transitions. The comparatively decreased intensity of pentacene transitions in the yellow sample suggests a slightly lower amount of pentacene.}
\zfl{SI_XRD_Comp}
\end{figure*}

The original paper reporting crystallographic information for the 2:1 pentacene cocrystal \cite{Mattheus2002}, and subsequent papers reporting cocrystal production during the sublimation of pentacene \cite{Roberson2005,Kim2021}, detail a distribution of colors in the cocrystal formation. Mattheus et al. describe the formation of cocrystals with color ranging "from dark-pink/red to red to white-transparent," but note that both colors have the same morphology and crystal structure \cite{Mattheus2002}. We find both the pink and yellow-white cocrystal products form during our PVT crystal growth. We hypothesize that the samples differ because the yellow cocrystal has trace amounts more of DHP located at the pentacene lattice sites, based on combined information from X-ray crystallography, solid-state diffuse-reflectance UV/Vis, and ODMR, which we detail here.

We performed single crystal X-ray diffraction on multiple pink and yellow needles for characterization. Structural analysis was performed with Mercury software, with average values reported with uncertainties in Table \ref{Tab:XRD_Analysis} (n=4 for the pink cocrystal samples, and n=3 for the yellow cocrystal). All reported distances are measured between the centroids of the respective molecules, except (1), (2), and (3) which are defined as the distance between two pentacene molecular planes, the distance between the molecular plane of one pentacene molecule and the centroid of another, and the distance between two carbon atom-centroids on nearest pentacenes, respectively. All angles are measured as the angle between molecular planes, except (6), which is measured as the angle between centroids and a third carbon atom kept consistent for reference. From this analysis, we find that the structures of either crystal do not have large differences in angle or distance either between pentacene molecules or between pentacene and DHP molecules in the crystal lattice. 

\begin{table*}[t]
\centering
\caption{\textbf{Crystal-structure analysis of pink and yellow 2:1 DHP/Pc cocrystals.}}
\label{Tab:XRD_Analysis}
\small
\renewcommand{\arraystretch}{1.35}
\begingroup
\arrayrulecolor{SITableRule}
\setlength{\arrayrulewidth}{0.45pt}
\begin{tabular}{|>{\raggedright\arraybackslash}P{8.2cm}|P{2.6cm}|P{2.5cm}|}
\hline
\rowcolor{SITableHeader}
\textbf{XRD parameter} & \textbf{Pink cocrystal} & \textbf{Yellow cocrystal} \\
\hline
(1) Nearest pentacene--pentacene distance along $a$ (\AA) & $6.13\pm0.004$ & $6.14\pm0.003$ \\
(2) Nearest pentacene--pentacene distance along $b$ (\AA) & $8.09\pm0.002$ & $8.08\pm0.005$ \\
(3) Nearest pentacene--pentacene distance along $c$ (\AA) & $4.10\pm0.005$ & $4.10\pm0.002$ \\
\hline
(4) Angle between nearest pentacenes along $a$ ($^\circ$) & $26.86\pm0.023$ & $26.81\pm0.037$ \\
(5) Angle between nearest pentacenes along $b$ ($^\circ$) & $51.48\pm0.049$ & $51.35\pm0.021$ \\
(6) Angle between nearest pentacenes along $c$ ($^\circ$) & $5.24\pm0.004$ & $5.26\pm0.005$ \\
\hline
(7) Nearest pentacene--DHP distance (\AA) & $5.23\pm0.001$ & $5.23\pm0.001$ \\
(8) Next-nearest pentacene--DHP distance (\AA) & $8.93\pm0.003$ & $8.94\pm0.003$ \\
(9) Nearest DHP--DHP distance (\AA) & $9.40\pm0.002$ & $9.41\pm0.003$ \\
\hline
(10) Angle between nearest pentacene and DHP ($^\circ$) & $51.82\pm0.050$ & $51.75\pm0.012$ \\
(11) Angle between next-nearest pentacene and DHP ($^\circ$) & $3.69\pm0.032$ & $3.78\pm0.037$ \\
(12) Angle between nearest DHP molecules ($^\circ$) & $52.47\pm0.056$ & $52.45\pm0.022$ \\
\hline
\end{tabular}
\endgroup
\end{table*}

All yellow cocrystal structures returned by the crystallographer were assigned $\mathrm{sp}^3$ ($-CH_2$) carbons at the central ring of pentacene without a corresponding tetrahedral molecular geometry like that of the central, tetrahedral $\mathrm{sp}^3$ carbons of DHP (compare \zfr{SI_XRD_Comp}A and B), while only half of the pink cocrystal structures submitted were returned with this structure.

While unusual, a planar $\mathrm{sp}^3$ carbon in a polycyclic aromatic hydrocarbon is possible. DHP-type molecules have been shown to adopt planar conformations in the solid state \cite{tajima_disproportionation-induced_2019} because the barrier between bent and planar ring conformations is $\app4$~kcal\,mol$^{-1}$. Additionally, this study shows that changes in DHP structure give rise to a range of fluorescence emissions (see Fig.~S4 of Ref.~\cite{tajima_disproportionation-induced_2019}).

Based on our experiments, we hypothesize that there is disorder in the crystal lattice site corresponding to the "planar DHP" position. For example, there may be some amount of DHP at this expected pentacene lattice site in the 2:1 cocrystal, which looks like an average of DHP and pentacene at this site from the ensemble-perspective of XRD. 

\begin{figure}[t]
  \centering
  {\includegraphics[width=0.98\columnwidth]{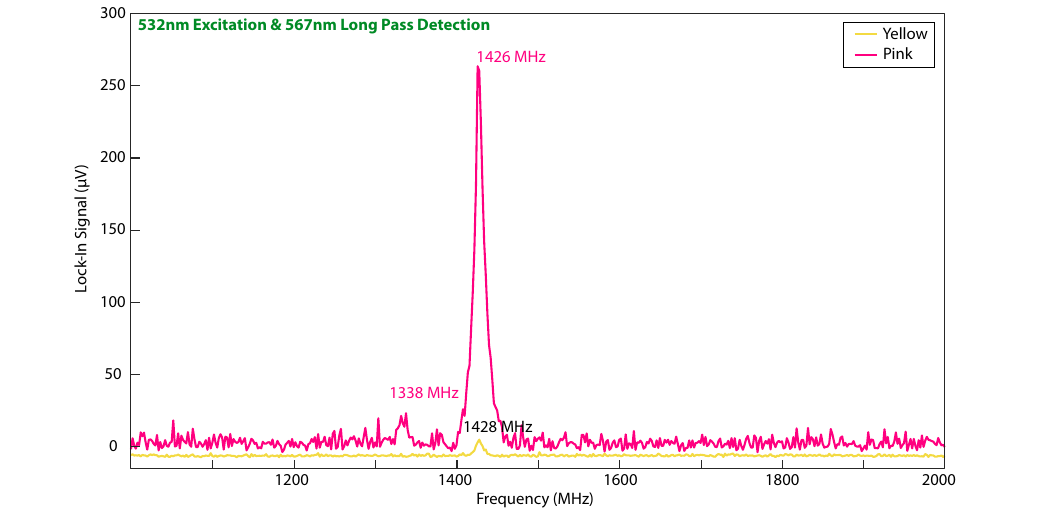}}
  \caption{\T{Cocrystal ODMR with 532~nm Excitation and 567~nm Detection} of \I{pink and yellow 2:1 6,13-DHP/Pc cocrystals}. Microwave power at the sample is -10 dBm. Intensity of the yellow cocrystal's $T_{\mathrm{XZ}}$ resonance is weaker than that of the pink cocrystal, suggesting lower pentacene concentration. }
\zfl{SI_ODMR_PinkYellow}
\end{figure}

The yellow and pink cocrystals have essentially identical wavelengths for all UV/Vis absorption features, within the minimum instrument resolution of 2~nm (\zfr{SI_XRD_Comp}C). We see that pentacene absorptions are retained in the 500-625~nm region in both materials. The most salient difference is that both samples have different relative intensities for peaks in the pentacene region, compared to features in the $<500$ nm region between samples (potentially DHP). We note that relative peak intensities can depend on sample composition in addition to sample packing and scattering in the measurement. The relative increase in intensity for $<$500~nm features and corresponding decrease in intensity of pentacene spectral features for the yellow cocrystal compared to the pink cocrystal sample is consistent with some fraction of DHP molecules replacing pentacene molecules in the yellow 2:1 cocrystal structure.

The pink and yellow cocrystals appear identical under the same 532~nm excitation / 567~nm detection conditions as measurements done in main paper \zfr{mfig3}, with decreased intensity for the yellow cocrystal resonances (\zfr{SI_ODMR_PinkYellow}). While the yellow sample's $T_{\mathrm{YZ}}$ resonance is below the noise floor, the observable reduction in $T_{\mathrm{XZ}}$ signal is consistent with reduced pentacene concentration in this sample as per our hypothesis. In summary, among the available cocrystals formed from pentacene sublimation, the pink cocrystal is the optimal choice for ODMR-based coherent control experiments under the measurement conditions we employ.

\section{SEM and Data Processing in ImageJ}
\label{SI_SEM}
3 to 4 cocrystal needles were mounted on SEM plates without further coating and imaged with a 0.7 kV electron beam and 100 pA probe-beam current in a Zeiss Crossbeam 550 SEM instrument at the Electron Microscope Lab using top-down, secondary electron-based detection. 

tiff files of all the SEM images were converted to .jpg and opened in the ImageJ-based software, Fiji. Replicate measurements of the SEM image scalebar were made using the Measure functionality and summarized to obtain an average value for the length in pixels of the scalebar. This conversion of pixels to length was used to set the image scale. With the scalebar set, the same procedure of replicate measurements summarized to produce an average value was used to determine the average width of a cocrystal needle at its center, shown in \zfr{mfig1}A. 

\section{Characteristic Inter-pentacene Spacing in 0.1\% Pentacene-doped \emph{p}-Terphenyl}
\label{SI_DistanceEstimation}
Lattice parameters of 0.1\% (w/w) pentacene-doped \emph{p}-terphenyl ($a=8.113$~\AA{}, $b=5.633$~\AA{}, and $c=13.600$~\AA{}) were obtained from Ref.~\cite{Ai_2023}, which reports a PDP unit-cell volume of $621.03~\text{\AA}^{3}$. There are two \emph{p}-terphenyl molecules per PDP unit cell, or $0.0032204~\text{\AA}^{-3}$. Thus, 0.1\% (w/w) (0.0826~mol\%) pentacene doping yields a pentacene number density of $n=2.643\times10^{-6}$ molecules~$\text{\AA}^{-3}$. A characteristic spacing can be defined as the reciprocal cube root of this number density, $n^{-1/3}=72.33$~\AA{}. This is a density-derived length scale, rather than a crystallographically determined nearest-neighbor or minimum pentacene--pentacene separation. Actual dopant separations depend on the substitutional distribution within the host lattice.

\begin{table*}[p]
\centering
\caption{\textbf{Room-temperature, optically detected ensemble $T_2$ coherence times across coherently controllable spin materials.} Spin-site concentrations and densities are reported in ppm and $\mathrm{cm}^{-3}$, respectively. For photoexcited molecular materials, these values denote triplet-forming molecular sites rather than the instantaneous photoexcited-triplet population. When a density was reported as a range or order of magnitude, the midpoint was used to estimate the concentration in ppm.}
\label{tab:funtable}
\footnotesize
\setlength{\tabcolsep}{4pt}
\renewcommand{\arraystretch}{1.25}
\begingroup
\arrayrulecolor{SITableRule}
\setlength{\arrayrulewidth}{0.45pt}
\begin{tabular}{|P{3.2cm}|P{2.7cm}|P{4.2cm}|P{1.7cm}|P{1.8cm}|}
\hline
\rowcolor{SITableHeader}
\textbf{Material} & \textbf{Spin-site concentration (ppm)} & \textbf{Spin-site density ($\mathrm{cm}^{-3}$)} & \textbf{$T_2$} & \textbf{Reference} \\
\hline
\rowcolor{SITableHighlight}
\textbf{2:1 DHP/Pc cocrystal} & $333{,}333$ & $9.514\times10^{20}$ & $0.874~\mu\mathrm{s}$ & \textbf{This work} \\
\hline
NV defects in diamond & $45$ & $7.933\times10^{18}$ & $0.44~\mu\mathrm{s}$ & \cite{Kucsko_2018} \\
\hline
NV defects in diamond & $1.70\times10^{-5}$ & $3\times10^{12}$ & $0.8~\mathrm{ms}$ & \cite{Bar-Gill_2013} \\
\hline
NV defects in SiC & $0.0387$ & $1.87\times10^{15}$ & $17.1~\mu\mathrm{s}$ & \cite{Wang_2020} \\
\hline
NV defects in SiC & $1.305$ & $10^{16}$ & $2.07~\mu\mathrm{s}$ & \cite{Jiang_2023} \\
\hline
PDP single crystal & $826$ & $2.660\times10^{18}$ & $2.7~\mu\mathrm{s}$ & \cite{Singh24} \\
\hline
PDP single crystal & $100$ & 0.01\% concentration; density not reported & $1.17~\mu\mathrm{s}$ & \cite{Mena24} \\
\hline
PDP single crystal & $1000$ & $3.3\times10^{18}$& $0.98~\mu\mathrm{s}$ & \cite{Huang_2026} \\
\hline
d-Pc:d-pT single crystal & $1000$ & $3.3\times10^{18}$ & $1.05~\mu\mathrm{s}$ & \cite{Huang_2026} \\
\hline
PDN single crystal & $400$ & $2.1\times10^{18}$ & $1.3~\mu\mathrm{s}$ & \cite{Huang_2026} \\
\hline
Pc:Picene single crystal & $1000$ & $2.9\times10^{18}$ & $1.51~\mu\mathrm{s}$ & \cite{Huang_2026} \\
\hline
PDP thin film & $1000$ & 0.1~mol\%; density not reported & $920~\mathrm{ns}$ & \cite{Mena24} \\
\hline
PDP 200~nm nanocrystals & $5000$ & 0.5\%; density not reported & $1.1~\mu\mathrm{s}$ & \cite{Ishiwata2025} \\
\hline
PDP 500~nm nanocrystals & $5000$ & 0.5\%; density not reported & $0.67~\mu\mathrm{s}$ & \cite{Ishiwata2025} \\
\hline
DAP:pT single crystal & $100$ & 0.01\% concentration; density not reported & $1.71~\mu\mathrm{s}$ & \cite{Mann_2025} \\
\hline
DAP:pT 447~nm nanocrystals & $1000$ & 0.1\%; density not reported & $1.46~\mu\mathrm{s}$ & \cite{Mann_2025} \\
\hline
Boron vacancies in h$^{10}$B$^{15}$N & $150$ & Density not reported & $186~\mathrm{ns}$ & \cite{Gong_2024} \\
\hline
Boron vacancies in hBN & $5.6$ & $5.4\times10^{17}$ & $2~\mu\mathrm{s}$ & \cite{Gottscholl_RT_2021} \\
\hline
Silicon vacancy in SiC & $0.00414$ & $4\times10^{14}$ & $8~\mu\mathrm{s}$ & \cite{Simin_2017} \\
\hline
Silicon vacancy in SiC & $0.00517$ & $5\times10^{14}$ & $105~\mu\mathrm{s}$ & \cite{Lekavicius_2022} \\
\hline
Silicon vacancy in SiC & $0.00403$ & $3.9\times10^{14}$ & $47~\mu\mathrm{s}$ & \cite{Kasper_2020} \\
\hline
Silicon vacancy in SiC & $0.0000515$ & $10^{12}$ & $0.85~\mathrm{ms}$ & \cite{Nagy_2019} \\
\hline
Silicon vacancy in SiC & $0.775$ & $7.5\times10^{16}$ & $7.8~\mu\mathrm{s}$ & \cite{wang23} \\
\hline
Silicon vacancy in SiC & $0.003625$ & $3.5\times10^{14}$ & $2.8~\mu\mathrm{s}$ & \cite{Stuermer_2026} \\
\hline
Neutral divacancies in SiC & $0.09$ & $(7\text{--}11)\times10^{15}$ & $5.6~\mu\mathrm{s}$ & \cite{Falk_2013} \\
\hline
\end{tabular}
\endgroup
\end{table*}

\section{Electron Concentration and Density Across Quantum Sensing Materials}
\label{SI_Density}
As tabulated in Table~\ref{tab:funtable}, the pentacene site density ($\mathrm{cm}^{-3}$) for the cocrystal was calculated by dividing two pentacene molecules by the unit-cell volume of $2102.12~\text{\AA}^{3}$ (given by Mercury after inputting XRD data), or $2.10212\times10^{-21}~\mathrm{cm}^{3}$, to yield a pentacene site density of $9.514\times10^{20}~\mathrm{cm}^{-3}$. This is the density of pentacene molecules capable of forming photoexcited triplets, rather than a measurement of the instantaneous triplet-spin population.

Similarly, for pentacene-doped \emph{p}-terphenyl, the number of pentacene molecules per unit-cell volume was determined by multiplying two \emph{p}-terphenyl molecules per PDP unit cell \cite{Singh2024_PT} by 0.0826~mol\% (0.1\% (w/w) doping) to give 0.00165 pentacene molecules per unit-cell volume. Dividing this value by a unit-cell volume of $621.03~\text{\AA}^{3}$ \cite{Ai_2023}, or $6.2103\times10^{-22}~\mathrm{cm}^{3}$, gives a pentacene site density of $2.660\times10^{18}~\mathrm{cm}^{-3}$.

NV defect centers in diamond have a diamond lattice parameter $a=3.567$~\AA{} \cite{Fahy_1987}, which corresponds to a cell volume of $45.38~\text{\AA}^{3}$, or $45.38\times10^{-24}~\mathrm{cm}^{3}$. Eight carbon atoms per unit-cell volume of $45.38\times10^{-24}~\mathrm{cm}^{3}$ gives $1.763\times10^{23}$ C atoms~$\mathrm{cm}^{-3}$. Multiplying this number by an NV-center doping of 45 NV centers per $10^6$ C atoms \cite{Kucsko_2018} yields an NV spin-site density of $7.93\times10^{18}~\mathrm{cm}^{-3}$. Both PDP crystals and NV-defect-center platforms exhibit spin-site densities two orders of magnitude lower than the pentacene site density of the cocrystal in this work.

Lattice parameters for the 4H-SiC unit cell ($a=3.0730$~\AA{} and $c=10.053$~\AA{}) and the 6H-SiC unit cell ($a=3.0806$~\AA{} and $c=15.1173$~\AA{}) \cite{Madelung_1982} are used to find a unit-cell volume of $8.2\times10^{-23}~\mathrm{cm}^{3}$ for the 4H-SiC unit cell and $1.2\times10^{-22}~\mathrm{cm}^{3}$ for the 6H-SiC unit cell. For the 4H-SiC unit cell, a Si-atom density is then calculated by dividing eight atoms in a 4H-SiC unit cell by $8.2\times10^{-23}~\mathrm{cm}^{3}$ to obtain $9.7\times10^{22}~\mathrm{cm}^{-3}$. The spin-site concentration is calculated by dividing the Si-vacancy density by $9.7\times10^{22}~\mathrm{cm}^{-3}$. Following a similar process, a Si-atom density is calculated by dividing twelve atoms in a 6H-SiC unit cell by $1.2\times10^{-22}~\mathrm{cm}^{3}$ to obtain $1\times10^{23}~\mathrm{cm}^{-3}$. The spin-site concentration is then calculated by dividing the Si-vacancy density by $1\times10^{23}~\mathrm{cm}^{-3}$. All concentrations are converted to ppm by multiplying by 10$^6$.

For the hBN defects, a unit-cell volume of $4.14\times10^{-23}~\mathrm{cm}^{3}$ was calculated from the reported lattice parameters \cite{Pease_1952}. 4 atoms per hBN unit cell were divided by $4.14\times10^{-23}~\mathrm{cm}^{3}$ to obtain $9.64\times10^{22}$ atoms per cubic cm. The reported density of boron vacancies is then divided by $9.64\times10^{22}$ (and multiplied by 10$^6$) to obtain the corresponding spin-site concentration.
\section{Solid-State Spectrofluorometry}
\label{SI_Spectrofluorometry}
Cocrystal samples were placed in a black, polystyrene 96-well plate. A 300-800~nm fluorescence intensity scan was performed using a Tecan Spark plate reader with 532~nm excitation wavelength and top fluorescence detection. A blank well was scanned and used for background subtraction. Results for the pink cocrystal are shown in \zfr{Fl}. For a 532~nm excitation, the sample has a fluorescence maximum at 652~nm.

\begin{figure}[t]
    \centering
    \includegraphics[width=0.90\columnwidth]{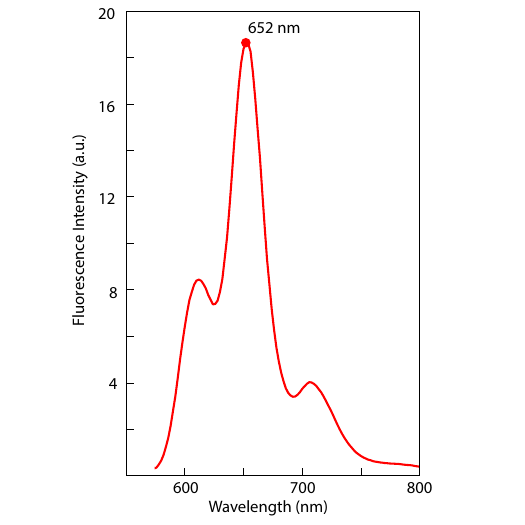}
\caption{\T{Cocrystal Fluorescence} Pink cocrystal fluorescence with 532 nm excitation has the strongest emission at 652 nm.}
    \label{Fluorescence}
    \zfl{Fl}
\end{figure}

\section{Solid-State Diffuse Reflectance UV/Vis}
\label{SI_SS_UVVis}
$\app30$~mg of pink cocrystal and 0.1\% (w/w) PDP crystals were powdered in a mortar and pestle and packed into a 3D-printed rectangular sample well. Diffuse-reflectance solid-state UV/Vis spectra of the samples were obtained using a Shimadzu UV 2600 instrument with a medium scan speed. Data were converted to pseudo-absorbance using the Kubelka--Munk transform, assuming a scattering coefficient of $\app1$, and plotted in \zfr{mfig1}D.

\section{Solution-State UV/Vis}
\label{SI_Soln_UVVis}
\begin{figure}[t]
    \centering
    \includegraphics[width=0.98\columnwidth]{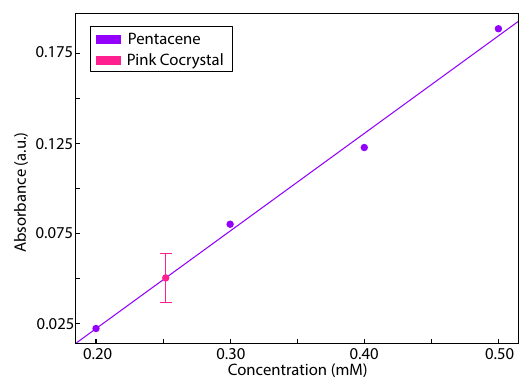}
    \caption{An absorbance (a.u.) versus concentration (mM) plot based on the 576~nm pentacene solution-phase absorption peak shows increasing absorbance for pentacene samples of increasing concentration in toluene, yielding a linear calibration curve y = 0.543x - 0.086 with an $R^2$=0.994. From this fit, a 0.926 mM pink cocrystal in toluene sample was determined to have a pentacene concentration of 0.2520$\pm$0.0254 mM. }
    \label{solnstateuvvis}
\end{figure}

Samples of varying concentration (0.5, 0.4, 0.3, and 0.2 mM) of pentacene dissolved in toluene were prepared alongside a 0.926~mM sample of pink cocrystal dissolved in toluene. A Varian Cary 50 UV/Vis spectrophotometer was blanked using a pure toluene sample and baseline-calibrated by blocking the beam. Samples were then loaded in a quartz cuvette, capped, and analyzed from 300--800~nm using a medium scan speed. The calibration curve is plotted in Fig.~\ref{solnstateuvvis} as a linear equation of best fit $y=0.543\pm0.032x-0.086\pm0.011$, and a corresponding $R^2=0.994$. The pink cocrystal's pentacene concentration is estimated from the fit to be $0.2520\pm0.0254$~mM, corresponding to $27.19\pm2.85$~mol\% pentacene in the cocrystal.

We note that the calculated pentacene concentration deviates from the expected 33.3~mol\%, which we hypothesize is due to electronic interactions of pentacene in solution. Pentacene has poor solubility in many solvents, and slight solubility in toluene. Prepared toluene calibration standards were included in the calibration curve if pentacene did not appear to be visibly aggregating in the cuvette, yet lower energy spectral features corresponding to electronically coupled pentacene molecules were still visible at 680~nm for even the 0.2mM standard \cite{ringstrom_molecular_2022}. These spectral features indicate that concentration-dependent intermolecular electronic interactions of pentacene are systematically affecting all standards used for this calibration curve, to which we attribute the discrepancies between calculated and theoretical pentacene mol \%.

\section{EPR and EasySpin Fitting}
\label{SI_EPR}
DHP/Pc cocrystal was ground into a powder with a mortar and pestle and loaded into a 3~mm borosilicate EPR tube to a height of approximately $2$~mm. The sample tube was loaded into a home-built transient EPR spectrometer employing IQ homodyne detection. The tube was then inserted into a loop-gap resonator (Bridge 12 B12T XLP resonator, loaded Q of 680 with sample) with a coupling loop whose height could be adjusted to change the coupling. A signal generator (ThinkRF LS1291D) supplied a continuous input microwave (-20 dBm, 0.01 mW, 9.565 GHz) into the resonator through a circulator. The sample within the resonator was photoexcited with 100~ns, 2.5 mJ pulses of 527 nm laser light (Photonics DS-527-25 pulsed Nd:YLF, with a depolarizing filter placed at the output). The trEPR signal after photoexcitation goes through the detection arm of the circulator where it is amplified by two low-noise amplifiers (LNA) in a chain, with the first LNA being a Qorvo CMD319 evaluation board. Microwave isolators were placed in the detection arm to reduce back-reflections of the microwave signal. The signal passes through a bandpass filter for 9.5 GHz before entering a passive IQ mixer for detection. The downconverted trEPR signal goes through low pass filtering and a DC block before being amplified by a DC amplifier (two cascaded LT1028 amplifiers with a total gain of 32 dB) and passed into an oscilloscope (Tektronix DPO3014). The oscilloscope records trEPR spectra in the time domain, triggered by the laser pulse using a fast photodiode. The time domain signal is integrated from 1.5 to 3.4 microseconds and plotted for each magnetic field to produce \zfr{mfig2}A. The trEPR signal corresponding to the low-field peak near 315 mT in the powder spectrum did not show signs of lifetime change due to power saturation between -20 dBm and -15 dBm, and so it was assumed at -20 dBm input microwave power no saturation was occurring for the spectrum. The electromagnet was a GMW 3472-70. We note that the laser excitation wavelengths differ by 5 nm between the ODMR and EPR experiments, yet we presume that this has a negligible effect on the experimental data based on similar pentacene absorptions at both wavelengths (\zfr{mfig1}D). Data was signal averaged for 50 scans. 

Least squares minimization of the powder EPR spectrum shown was performed using the pepper function for powder spectra in MATLAB's EasySpin software package \cite{Stoll2006} to extract triplet state parameters. The fit was run until converged, that is when the RMSD was minimized and the best fit parameters were self-consistent with the initial guess parameters. Best fit parameters are tabulated in Table \ref{tab:EasySpinFit} with uncertainties; population values are re-normalized to sum to one. 

\begin{table*}[t]
\centering
\caption{\textbf{EasySpin best-fit parameters for the 2:1 6,13-dihydropentacene:pentacene cocrystal.}}
\label{tab:EasySpinFit}
\small
\renewcommand{\arraystretch}{1.35}
\begingroup
\arrayrulecolor{SITableRule}
\setlength{\arrayrulewidth}{0.45pt}
\begin{tabular}{|P{4.2cm}|P{3.0cm}|P{3.0cm}|P{3.0cm}|}
\hline
\rowcolor{SITableHeader}
\textbf{EPR fit parameter} & \textbf{Value} & \textbf{Lower bound} & \textbf{Upper bound} \\
\hline
$g$ & $2.00001$ & $1.99972$ & $2.0003$ \\
\hline
Linewidth (lw) & $1.45534$ & $1.24807$ & $1.66261$ \\
\hline
$D$ (MHz) & $1376.8$ & $1371.4$ & $1382.2$ \\
\hline
$E$ (MHz) & $-47.4$ & $-49.5$ & $-45.2$ \\
\hline
$p_x$ & $0.752$ & $0.735$ & $0.770$ \\
\hline
$p_y$ & $0.199$ & $0.191$ & $0.208$ \\
\hline
$p_z$ & $0.049$ & $0.040$ & $0.058$ \\
\hline
\rowcolor{SITableHighlight}
\textbf{XZ polarization} & $0.70$ & $0.695$ & $0.712$ \\
\hline
$RMSD$ & \multicolumn{3}{|c|}{$1.54$} \\
\hline
\end{tabular}
\endgroup
\end{table*}

\section{Solid-State FTIR}
\label{SI_SS_FTIR}
\begin{figure*}[t]
    \centering
    \includegraphics[width=0.88\textwidth]{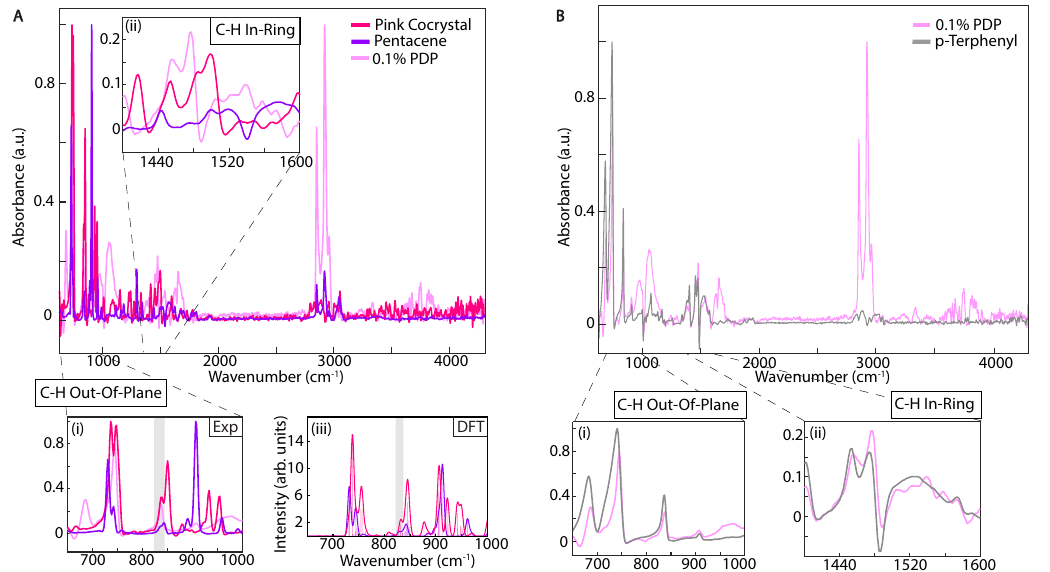}
    \caption{\T{Solid-state FTIR spectra.} (a) 2:1 dihydropentacene (DHP):pentacene pink cocrystal (dark pink), 0.1\% (w/w) PDP (light pink), and pentacene (purple) FTIR spectra are superimposed for comparison. Aromatic out-of-plane (oop) and aromatic in-ring C-H vibrations are emphasized in insets (i) and (ii), respectively. Inset (iii) shows the DFT-simulated vibration spectrum in the same C-H oop region as inset (i) with enhanced 837~$cm^{-1}$ peak highlighted. (b) 0.1\% (w/w) PDP (light pink) and p-terphenyl (grey) FTIR spectra are similarly compared. Aromatic out-of-plane and aromatic in-ring C-H vibrations are emphasized in insets (i) and (ii), respectively. The PDP vibrational structure is dominantly that of the p-terphenyl host material.}
    \zfl{SS_FTIR}
    \label{SS_FTIR}
\end{figure*}

\begin{figure*}[t]
    \centering
    \includegraphics[width=0.88\textwidth]{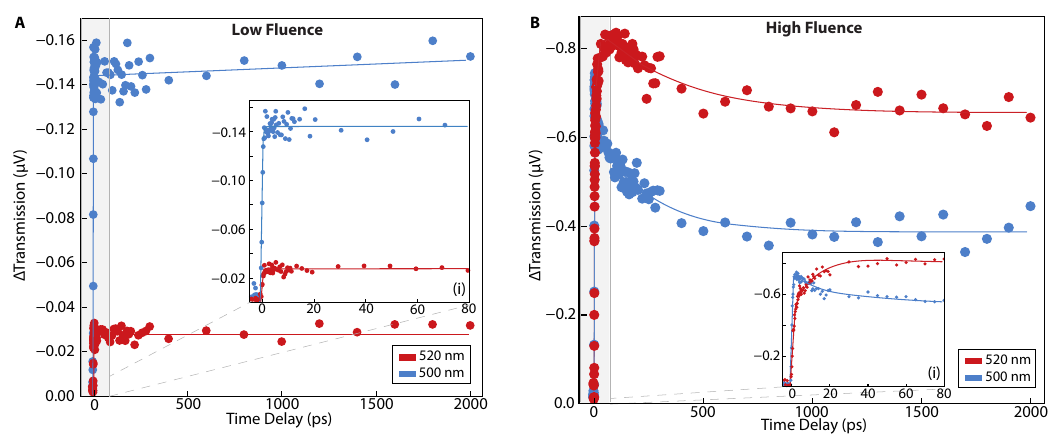}
\caption{\T{TAM Measurements of a DHP/Pc Cocrystal} performed at (A) \I{low} (62 $\mu$J/cm\textsuperscript{2}) and (B) \I{high} (226 $\mu$J/cm\textsuperscript{2}) pump fluences. The inset (i) zooms in on time delays from 0-80 ps. Under low fluence conditions, which do not damage the crystal, the signal amplitude does not decay over the entire available time window (2 ns), indicating negligible triplet formation. The lack of triplet formation during this SF-relevant time window indicates that SF is likely not active within the cocrystal. At high fluence, ESA probed at 520 nm grows with time, indicating the formation of triplet excitons. This growth is accompanied by a corresponding loss of ESA signal at 500 nm from singlet excitons. These observations are consistent with rapid triplet formation via SF. We attribute this change in dynamics to partial melting of the crystal, which allows pentacene molecules to form structures conducive to SF. Solid lines are fits to the data using a convolution of multiple exponentials with a step function representing the finite instrument response of the TAM.}
  \zfl{SI_TAM}
  \label{SI_TAM}
\end{figure*}

A ZnSe trough plate and ATR-FTIR PikeATRMax II accessory was set up on a Bruker Vertex 70 FTIR instrument. The sample chamber was purged with nitrogen gas for 5~min before collecting a background spectrum. Powdered sample was loaded onto the crystal in a line and compressed with the accessory. Data were acquired from $400$--$4500~\mathrm{cm}^{-1}$ with 100 scans per dataset. The instrument resolution was set to $4~\mathrm{cm}^{-1}$, and the angle of incidence was approximately $45^\circ$. All data were ATR-corrected to absorbance in the instrument software. Subsequently, all data were baseline-corrected using the \I{msbackadj} MATLAB function with a window size of 100 and a step size of 50. The data presented are the average of three replicate measurements on each sample. This procedure was repeated for cocrystal, pentacene, and 0.1\% pentacene-doped \emph{p}-terphenyl samples. Data are shown in Fig.~\ref{SS_FTIR}. DFT-simulated FTIR data are shown in inset (iii) alongside the C-H oop experimental data to highlight an enhanced amplitude vibration at 837~$cm^{-1}$, which is the 807~$cm^{-1}$ $T_X$-populating vibrational mode identified in Ref. \cite{Sakamoto2023} mapped to the experimental data.

\begin{figure}[t]
    \centering
    \includegraphics[width=\columnwidth]{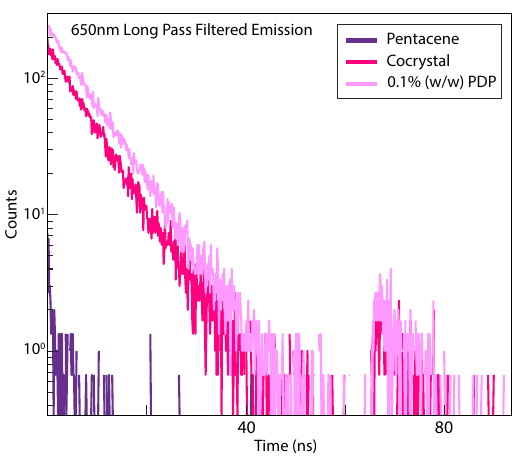}
\caption{\T{Representative TCSPC datasets} showing comparable fluorescence lifetimes between the dense 2:1 cocrystal (pink) and the dilute doped crystal system of 0.1\% (w/w) PDP (light pink). Both of these samples have appreciable fluorescence lifetimes, compared to that of the rapid SF material, pure pentacene (purple).}
  \zfl{SI_TCSPC}
  \label{SI_TCSPC}
\end{figure}

\section{Transient Absorption Microscopy (TAM)}
\label{sec:SI_TAM}
TAM measurements were performed on a home-built setup. A 20 W, 532~nm cw Nd:YVO$_4$ laser (Spectra-Physics, Millennia eV) was split 70:30 to pump a 200 kHz Ti:Sapphire regenerative amplifier (Coherent, RegA 9000, 14 W) and its seed oscillator (Coherent, Mira 900-F, 6 W). Part of the amplifier's 800 nm output was converted to a 600 nm pump via an OPA (Coherent, OPA Model 9400), while the remainder was focused into a sapphire crystal to generate a broadband probe that was spectrally filtered to either 500 nm using a Thorlabs FBH500-10 filter (10 nm bandwidth) or 520 nm using a Thorlabs FBH550-10 filter rotated to shift the center wavelength (\zfr{SI_TAM}). The pump was focused through a 50~$\mu$m pinhole and a double-pass delay stage (2 ns scan range), and was modulated to 100 kHz with an acousto-optic modulator (Gooch \& Housego, 3100-125). Pump and probe pulses were combined collinearly and focused onto a cocrystal needle of appropriate thickness for transmission measurements using a 150X, 0.95 NA air objective, giving a pump beam waist diameter of $\app 0.58~\mu$m ($1/e^2$). The beam waist was determined by scanning a knife edge across the focal point. Transmitted light was collimated by a 20X, 0.4 NA objective (Edmund Optics, DIN 20/0.4), filtered to block residual pump (530 nm notch, 40 nm bandwidth), and detected on an avalanche photodiode (Menlo Systems, APD210) coupled to a lock-in amplifier (Signal Recovery, 7280 DSP) referenced to the 21st harmonic of the pump modulation frequency (time constant 10 ms, sensitivity 5~$\mu$V). The IRF had a $\app1$ ps FWHM determined from fits to the data. Pump and probe polarizations were oriented parallel to one another and measurements were performed at room temperature in ambient atmosphere. Each time point was averaged over 200 lock-in readings, and 10 scans were averaged in total for each dataset.

Low- and high-fluence measurements were taken using pump fluences of 62 and 226~$\mu$J/cm², respectively. As discussed in the main text, at low fluence, we see an instrument-limited rise of excited state absorption (ESA) stemming from singlet excitons. This signal persists without change over our measurement time window (\zfr{SI_TAM}A), indicating that triplet excitons in the cocrystal are produced over timescales longer than a few nanoseconds. This slow production of triplet excitons is consistent with their generation via intersystem crossing rather than singlet fission (SF). 

In contrast, at high fluence we observe different behavior (\zfr{SI_TAM}B). Specifically, the ESA band monitored at 500 nm is found to decay over time concomitant with a rise in ESA signal seen at 520 nm. These dynamics are consistent with what is expected for active SF in pentacene, wherein an overlapping triplet ESA signal should appear rapidly near 520 nm while the singlet ESA near 500 nm should decay on the same timescale as singlet excitons are converted into triplet pairs\cite{lee_two_2019,bender_surface_2018,pensack_solution-processable_2017}. This change in dynamics is accompanied with a photoinduced change in the cocrystal’s optical properties, suggesting that the cocrystal’s structure is altered within the pump volume by the power contained in the pump. We hypothesize that these changes result from partial melting of the crystal, which can allow pentacene molecules to associate into geometries that are conducive to SF. The absence of these dynamics in the low fluence measurements, in which no apparent photoinduced alterations in the crystal were observed, suggests that the cocrystal’s native packing structure is not conducive to SF.

\section{Time-Correlated Single Photon Counting (TCSPC)}
\label{sec:SI_TCSPC}
A microscope coverslip was covered with black tape. NOA81 adhesive was applied to the surface, and 3--4 crystals were placed on it. The samples were UV-cured until the adhesive solidified and then inserted into a home-built TCSPC setup. Mode-locked 800~nm laser pulses at a repetition rate of 263~kHz were generated using a Coherent RegA 9050 regenerative amplifier seeded by a Coherent Vitara-T Ti oscillator. A visible-light continuum was produced by focusing the amplified pulses into a 1~mm-thick sapphire crystal. The continuum was first filtered using a Thorlabs FESH0700 700~nm short-pass filter, and the excitation wavelength was selected to match the sample absorption using a 520~nm bandpass filter with a full width at half-maximum (FWHM) of 40~nm (Thorlabs FBH520-40). The excitation power was adjusted to $30~\mu\mathrm{W}$ at the sample position, with an approximate excitation-beam diameter of $100~\mu\mathrm{m}$. Fluorescence emission was spectrally selected at the desired detection wavelengths using combinations of bandpass and long-pass filters. Emission above 650~nm was collected using a Thorlabs FELH0650 650~nm long-pass filter. The filtered fluorescence was detected using a Quantum Opus SN082 SNSPD superconducting nanowire single-photon detector system. Replicate measurements were taken on multiple samples, where $n>1$ as indicated, and averaged to obtain a dataset for an individual sample. A linear function was fit to the averaged data plotted as $\log(\mathrm{intensity})$ versus time to extract the fluorescence lifetime. Measurements on multiple samples were averaged to obtain the mean fluorescence lifetime of a material and its standard deviation.

Within the instrument timing resolution of $\app100~\mathrm{ps}$, there were insufficient fluorescence counts from a pure pentacene crystal to extract a fluorescence lifetime, as expected for a highly efficient singlet-fission material with an $\app80~\mathrm{fs}$ timescale \cite{Wilson_2011}. Only one sample run had appreciable counts, as reported in Fig.~\ref{SI_TCSPC}. Sufficient counts were obtained to fit the fluorescence lifetimes of the cocrystal and 0.1\% (w/w) PDP samples, which were comparable: $\tau=7.6\pm0.6~\mathrm{ns}$ ($n=3$) and $\tau=7.5\pm0.1~\mathrm{ns}$ ($n=3$), respectively. This suggests that singlet fission is not appreciably enhanced in the cocrystal despite the dense packing of pentacene, and that its behavior is comparable to that of 0.1\% (w/w) PDP. These conclusions agree with those from TAM.

\section{Nuclear Magnetic Resonance (NMR)}
\label{SI_NMR}
Samples containing pentacenequinone and the pink cocrystal at approximate concentrations of 2--3~mM in deuterated chloroform were each prepared in Norell 5~mm NMR tubes. $^1$H NMR spectra of each sample were recorded on a Bruker AVQ-400~MHz spectrometer, and peak assignments are summarized here. In \zfr{yaynmr}b, the peak at 4~ppm corresponds to the aliphatic dihydropentacene hydrogens, whereas the peaks from 7 to 8~ppm correspond to aromatic DHP and pentacene hydrogens. In \zfr{yaynmr}a, the peak at 8~ppm corresponds to aromatic hydrogens within three bonds of the pentacenequinone ketone and is thus characteristic of pentacenequinone. The absence of this resonance in \zfr{yaynmr}b indicates that pentacenequinone is not detected in the pink cocrystal, within the sensitivity of the measurement.

\begin{figure*}[t]
    \centering
    \includegraphics[width=0.88\textwidth]{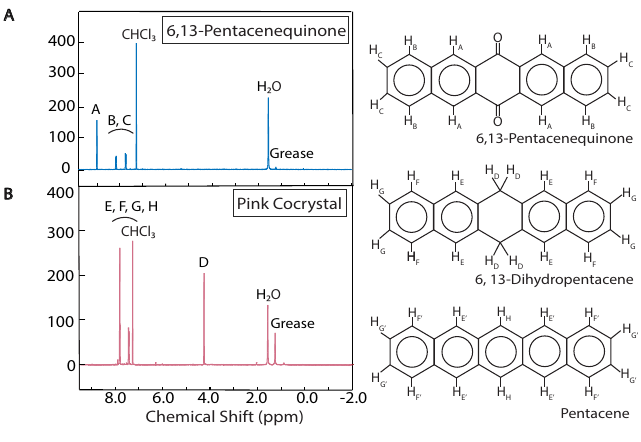}
\caption{\T{$^1$H NMR confirms dihydropentacene in the cocrystal.} $^1$H NMR spectra in $\mathrm{CDCl_3}$ of pentacenequinone and the pink DHP:pentacene cocrystal are recorded. Spectra are referenced to residual $\mathrm{CHCl_3}$ at 7.3~ppm; water and grease impurities appear at 1.6~ppm and 1.3~ppm, respectively \cite{Gottlieb_1997}. The resonance near 4~ppm corresponds to aliphatic DHP protons, while the 7--8~ppm region contains aromatic DHP and pentacene signals.}
    \zfl{yaynmr}
\end{figure*}

\section{Optically Detected Magnetic Resonance (ODMR)}\label{SI_ODMR}
\subsection{Experimental Setup}
\label{SI_ODMR_Setup}
The ODMR setup is comprised of two components: optical excitation and detection, as well as microwaves (MW) \cite{Singh24}. A continuous wave Coherent 532 nm laser is directed to the sample via a series of mirrors up to a 10X Olympus microscope objective, which focuses the beam onto the sample. Red sample photoluminescence (PL) is filtered through a dichroic mirror collinear with the laser excitation pathway and directed onto a Hamamatsu C12703 DC-10MHz avalanche photodiode detector (APD). The APD output is passed to a Stanford Research Systems SR830 DSP lock-in amplifier set to 1 kHz reference frequency, where the signal is recorded with 100 mV sensitivity.

MW are generated by a Tabor Proteus P9484M arbitrary waveform transceiver, then passed through a TTL-triggered Mini-Circuits ZASWA-2-50DRA+ switch gated on/off at 1 kHz using a Swabian Instruments PulseStreamer 8/2, which modulates the MW at the lock-in reference frequency. The MW are passed through two amplifiers, directed through a power meter, and applied to the sample via a two-turn Helmholtz coil, which functions as the resonator. The amplifier chain is a Mini-Circuits ZKL-33ULN-S+ and ZHL-2W-63-S+ for the high frequency (YZ, XZ) ODMR transitions and Mini-Circuits ZFL-500LN+ and LZY-22+ for the low-frequency XY transition.

\subsection{Sample Mounting for ODMR}
\label{SI_SampleMounting}
Each cocrystal sample was mounted inside a capillary tube that was sealed at one end with a UV-cured NOA81 adhesive. In the exposed end of the capillary tube, a single needle of the cocrystal was centrally positioned parallel to the setup's MW coil using the following procedure. A conservative amount of NOA81 adhesive is placed at the tip of a copper wire to which a pink cocrystal needle is vertically adhered. The non-sample end of the wire is inserted into a Signatone Model S-725-CLM micropositioner. The micropositioner is used to centrally position the cocrystal needle in the uncured adhesive end of the capillary tube, aided by a top-down viewing digital microscope. UV exposure cures the needle in place, and the copper wire is removed before a final UV-lamp curing.

\subsection{ODMR Linewidth Dependence on Microwave Power}
\label{SI_Linewidth}
ODMR spectra were taken at fixed optical power $\app30$ mW on the sample and variable microwave power to determine the power dependence of each triplet transition's linewidth shown in \zfr{SI_MWDependence}. MW-power unbroadened linewidths at the stated optical power are as follows for XZ, XY, and YZ transitions, respectively, reported with 95\% confidence intervals extracted from the fit: $7.55\pm0.20$~MHz, $10.19\pm0.60$~MHz, and $6.68\pm0.57$~MHz. Although the intercept is smaller for the YZ transition, the low SNR of this transition increased the uncertainty associated with this value and would make it suboptimal for sensing experiments (see \zfr{mfig3} of the main paper). 

\begin{figure*}[t]
    \centering
    \includegraphics[width=0.92\textwidth]{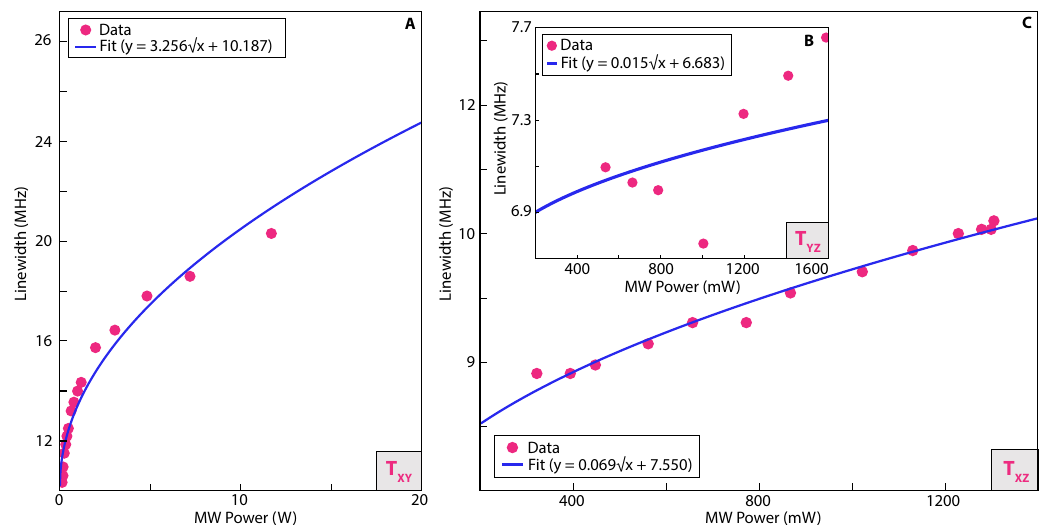}
\caption{\T{Linewidth dependence on microwave power} for the (A) $T_{\mathrm{XY}}$, (B) $T_{\mathrm{YZ}}$, and (C) $T_{\mathrm{XZ}}$ transitions. Data are fit to the function $l(P_{\mathrm{MW}})=l_0+a\sqrt{P_{\mathrm{MW}}}$. Y intercept, $l_0$, represents the intrinsic, power-unbroadened linewidth. Determination of the YZ linewidth power dependence is limited by the low SNR of the YZ transition.}
    \zfl{SI_MWDependence}
\end{figure*}

\section{Coherence Measurements}\label{SI_Coherence}
All spin-coherence measurements were conducted on $T_{\mathrm{YZ}}$ with a parallel ODMR microscopy setup, which we detail here. A 2~mW, 532~nm optical excitation beam was focused into an Olympus MXPLFLN 50$\times$ objective to spread the beam laterally and minimize damage from intense, focused excitation. A $20~\mu\mathrm{s}$ laser-polarization pulse was used for all coherence measurements except the spin-echo and CPMG measurements, which used a $5~\mu\mathrm{s}$ laser pulse to reduce triplet formation and improve echo and CPMG contrast. Microwave sequences specific to the coherence measurement were initiated after an $\app2~\mu\mathrm{s}$ gap following the laser-polarization pulse. A 1.6~mm-diameter, 0.51~mm-thick, 50~$\Omega$ impedance-matched microwave loop resonator was used for microwave delivery. A 30~ns amplifier-settling time was used for phase-sensitive nanosecond microwave-rotation pulses.

Sample emission was collected through the hole in the microwave resonator and recorded using a DET100A photodiode sampled via a SR860 lock-in amplifier or amplified through a preamplifier onto a National Instruments analog-to-digital converter (ADC). Coherence times were obtained by fitting the optical readout as a function of evolution time. 

\begin{figure}[t]
    \centering
    \includegraphics[width=0.98\columnwidth]{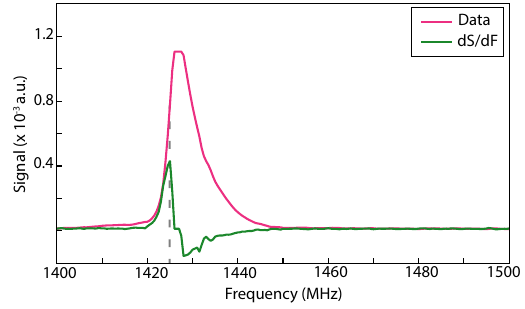}
\caption{\T{Dataset used to calculate maximum sensitivity} is shown overlaid with the numerical derivative. Dashed line is a guide to the eye for the frequency at which the derivative is a maximum, here 1425 MHz. Optical power at the sample was approximately 30 mW and MW power was approximately 480 mW on resonance.
$T_{\mathrm{XZ}}$ resonance is at $\approx$1428 MHz.}
    \zfl{SI_BestSensitivity}
    \label{SI_BestSensitivity}
\end{figure}

\section{DC Sensitivity Estimate}
\label{SI_Sensitivity}
The DC field sensitivity was calculated following the method employed in Ref.~\cite{Singh24} using
\begin{equation}
    \eta_{\mathrm{DC}}
    =\frac{\sigma\sqrt{\tau}}
    {\left(\dfrac{\mathrm{d}S}{\mathrm{d}F}\right)\gamma_{\mathrm{e}}}.
    \label{eq:DC_sensitivity}
\end{equation}
Here, $\sigma$ is the mean absolute deviation of the off-resonance signal (in this case from the 1460--1500~MHz window), $\tau$ is the integration time (for which we used the low-pass lock-in-amplifier settling time, $4\times300$~ms), $\mathrm{d}S/\mathrm{d}F$ is the maximum slope in the ODMR signal, and $\gamma_{\mathrm{e}}$ is the electron gyromagnetic ratio (28024.951 MHz/T). The best sensitivity observed from our measurements thus far is $157~\mathrm{nT}/\sqrt{\mathrm{Hz}}$ from the dataset in Fig.~\ref{SI_BestSensitivity}, which can be further optimized by improving ODMR excitation, with a pulsed laser, and detection by using a focusing objective to collect emitted light from the full sample field of view.

\section{Density Functional Theory (DFT) Energy Level Calculations}
\label{SI_DFT}
We use DFT energy-level calculations on multiple pentacene-containing crystal systems to probe ODMR-relevant photophysical processes. DFT calculations were performed in VASP using a plane-wave basis set and the projector-augmented-wave (PAW) method to describe the electron--ion interaction. Exchange--correlation effects were treated within the Perdew--Burke--Ernzerhof (PBE) generalized-gradient approximation (GGA). The Brillouin zone was sampled at the gamma point. Spin polarization was included, and orbital constraints were applied to maintain the pentacene molecules in different states. Geometry optimization was carried out using the conjugate-gradient algorithm with VASP's default convergence criteria.

Rotated pentacene dimer structures (\zfr{mfig4}A) were constructed by first removing all DHP molecules from the geometry optimized cocrystal crystal structure, keeping the lattice constants fixed. Next, one of the pentacene molecules was rigidly rotated about its long axis in 30$^\circ$ increments, from 0$^\circ$ to 180$^\circ$. Electronic structure calculations of the rotated pentacene dimers were performed using the same DFT exchange correlation functional and convergence parameters as the cocrystal structures.

\begin{figure*}[t]
    \centering
    \includegraphics[width=0.96\textwidth]{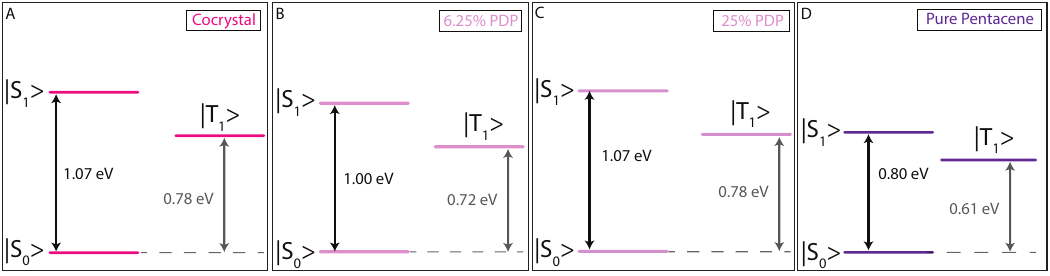}
\caption{\T{DFT energy-level comparisons for pentacene-doped \emph{p}-terphenyl.} DFT-calculated singlet/triplet energy-level diagrams for (A) the 2:1 DHP/Pc cocrystal for reference, (B) 6.25\% pentacene-doped \emph{p}-terphenyl (PDP), (C) 25\% PDP, and (D) pure pentacene.}
    \zfl{DFT_PDP}
    \label{DFT_PDP}
\end{figure*}

\begin{figure*}[t]
    \centering
    \includegraphics[width=0.68\textwidth]{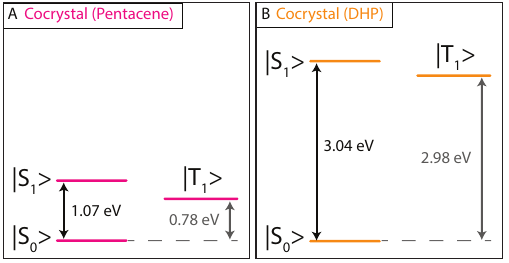}
\caption{\T{DFT triplet energy-level comparisons} for (A) pentacene in the DHP/Pc cocrystal and (B) a 6,13-dihydropentacene model in the cocrystal, constructed by removing the two pentacene molecules from the cocrystal unit cell while retaining the four 6,13-dihydropentacene molecules, their atomic positions, and the original lattice vectors. There is a significant $\approx2.2~\mathrm{eV}$ energy gap between a triplet localized on DHP and one localized on pentacene in the cocrystal structure, which blocks triplet migration despite potential orbital overlap between molecules.}
    \zfl{DFT_DHPPT}
    \label{DFT_DHPPT}
\end{figure*}

\begin{figure*}[t]
    \centering
    \includegraphics[width=0.88\textwidth]{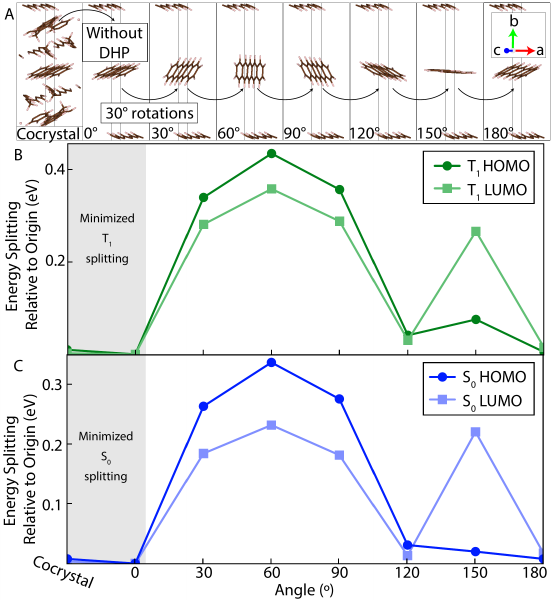}
\caption{\T{Angular Dependence of Orbital Energy Splitting in HOMO and LUMO for Vertical Pentacene Dimer Configurations} (A) \I{Unit-cell geometries varying angles of the vertical pentacene dimer} used for (B,C) \I{$T_1$ and $S_0$ orbital energy splitting vs. rotational angle DFT calculations}. The intrinsic cocrystal angle lies near a global minimum in splitting, consistent with weak intermolecular electronic coupling, rationalizing the retention of ODMR in the dense pentacene ensemble.}
    \zfl{DFT_Vertical}
    \label{DFT_Vertical}
\end{figure*}

\begin{figure*}[t]
    \centering
      \includegraphics[width=0.88\textwidth]{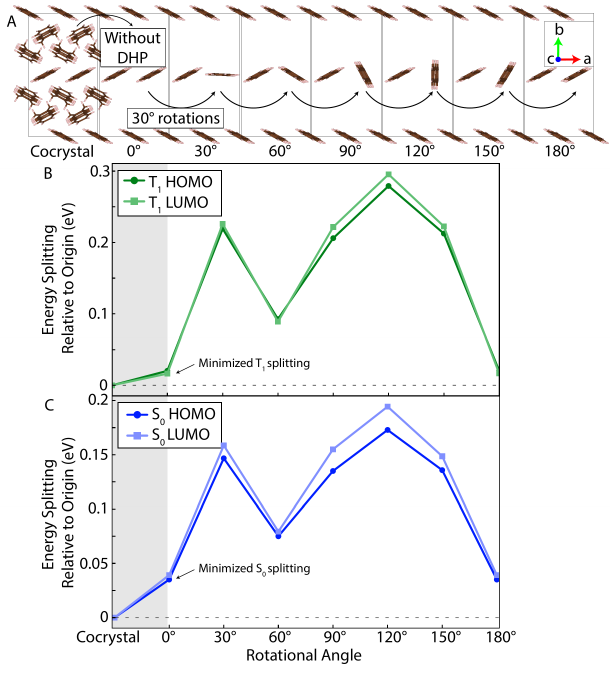}
\caption{\T{Angular Dependence of Orbital Energy Splitting in HOMO and LUMO for Short Axis, Horizontal Pentacene Dimer Configurations} (A) \I{Unit-cell geometries varying angles of the horizontal pentacene dimer along the in-plane molecular short axis} used for (B,C) \I{$T_1$ and $S_0$ orbital energy splitting vs. clockwise rotational angle DFT calculations}. The intrinsic cocrystal angle lies near a global splitting minimum for both electronic configurations, consistent with weak intermolecular electronic coupling, rationalizing the retention of ODMR in the dense pentacene ensemble.}
    \zfl{DFT_Horizontal}
    \label{DFT_Horizontal}
\end{figure*}

Calculations were set up for pure pentacene, which demonstrates TTA (Fig.~\ref{DFT_PDP}D); 6.25 mol \% PDP, close to the 0.1\% (w/w) which has previously been shown to exhibit room-temperature ODMR \cite{Singh24,Mena24} (Fig.~\ref{DFT_PDP}B); and 25\% PDP, which has a comparable volume density of pentacene molecules (Fig.~\ref{DFT_PDP}C). These are compared with the cocrystal Jablonski diagram (Fig.~\ref{DFT_PDP}A).
The 0.1\% (w/w) PDP sample was simulated at 6.25 mol\% doping level for computational efficiency: the simulation cell contained one pentacene molecule and 15 \emph{p}-terphenyl molecules, with lattice parameters $a = 16.212$~\AA, $b = 11.226$~\AA, $c = 27.226$~\AA, $\alpha = \gamma = 90^\circ$, and $\beta = 92.020^\circ$. This model corresponds to a pentacene mole fraction of 6.25\% calculated as
\[
x_{\mathrm{Pc}}
= \frac{N_{\mathrm{Pc}}}
       {N_{\mathrm{Pc}} + N_{\mathrm{p\text{-}terphenyl}}}
= \frac{1}{1+15}
= 0.0625,
\]
where $N_{\mathrm{Pc}}$ and $N_{\mathrm{p\text{-}terphenyl}}$
denote the numbers of pentacene and \emph{p}-terphenyl molecules in the simulation cell, respectively.
Similarly, the 25 mol\% PDP model was constructed using a periodic cell containing one pentacene molecule and three \emph{p}-terphenyl molecules, with lattice parameters
$a = 8.106$~\AA, $b = 11.226$~\AA, $c = 13.613$~\AA,
$\alpha = \gamma = 90^\circ$, and $\beta = 92.020^\circ$.
The pentacene mole fraction was therefore
$x_{\mathrm{Pc}} = 1/(1+3) = 0.25$.

The $S_0$, $S_1$, and $T_1$ states were modeled using constrained DFT with the orbital occupations specified below.
$S_0$: HOMO = $\uparrow\downarrow$, LUMO = empty. \\
$S_1$: HOMO = $\uparrow$, LUMO = $\downarrow$
(opposite spins; broken-symmetry singlet). \\
$T_1$: HOMO = $\uparrow$, LUMO = $\uparrow$
on one pentacene molecule. \\

We find that the $T_1$ energy is comparable for systems of similar pentacene volume density, such as the cocrystal and 25\% PDP.

SF has been observed in PDP films at pentacene concentrations above 0.95~mol\% \cite{Lubert-Perquel2018}; high molecular-site density might therefore suggest that it could compete in both a 25\% PDP model and the 33.3~mol\% cocrystal. We can understand this by comparing trends in DFT-calculated energy levels between samples, as a common energetic guideline for exoergic SF is $E(S_1)\gtrsim 2E(T_1)$, although SF rate and yield also depend on intermolecular geometry, electronic coupling, vibronic structure, and triplet-pair binding.

The present constrained-DFT energies give E($S_1$)=$1.07~\mathrm{eV}<2(E(T_1)=0.78~\mathrm{eV})=1.56~\mathrm{eV}$ for both the 25~mol\% PDP model and the cocrystal and E($S_1$)=$0.80~\mathrm{eV}<2(E(T_1)0.61~\mathrm{eV})=1.22~\mathrm{eV}$ for pure pentacene, indicating that exoergic SF is not favored within these calculations. Because this simple energetic test does not reproduce the reported behavior of high-concentration PDP films and pure pentacene, it cannot exclude SF by itself, and the cocrystal's triplet generation mechanism instead relies on the time-resolved spectroscopic evidence.

The optimized structures were used to calculate the force constants and Born effective charge tensors using VASP~\cite{Kresse1996}.
The resulting data were post-processed using the \texttt{dynmat.x} utility of Quantum
ESPRESSO~\cite{Giannozzi2009} to obtain $\Gamma$-point vibrational frequencies and infrared (IR) intensities within the harmonic approximation. 
The discrete IR transitions were broadened using Gaussian functions with a standard deviation of $\sigma = 3$~cm$^{-1}$, and the window corresponding to =C-H out-of-plane vibrations is compared with experimental data in Fig.~\ref{SS_FTIR}A(iii).

The spatial extent of each orbital in Fig.~\ref{DFT_Spread} was quantified along
the molecular long axis ($x$), in-plane short axis ($y$), and out-of-plane normal ($z$) using
\[
\sigma_{\alpha} =
\left[
\frac{\int_{\Omega}(r_{\alpha}-\bar{r}_{\alpha})^2
|\psi(\mathbf{r})|^2\,d\mathbf{r}}
{\int_{\Omega}|\psi(\mathbf{r})|^2\,d\mathbf{r}}
\right]^{1/2},
\qquad \alpha \in \{x,y,z\},
\]
where $r_{\alpha}$ is the coordinate along the corresponding molecular axis, $\bar{r}_{\alpha}$ is its orbital-density-weighted mean, and $\Omega$ denotes the spatial region included in the analysis.
A larger $\sigma_{\alpha}$ indicates a broader orbital-density distribution along that molecular axis.

\begin{figure*}[t]
    \centering
    \includegraphics[width=0.82\textwidth]{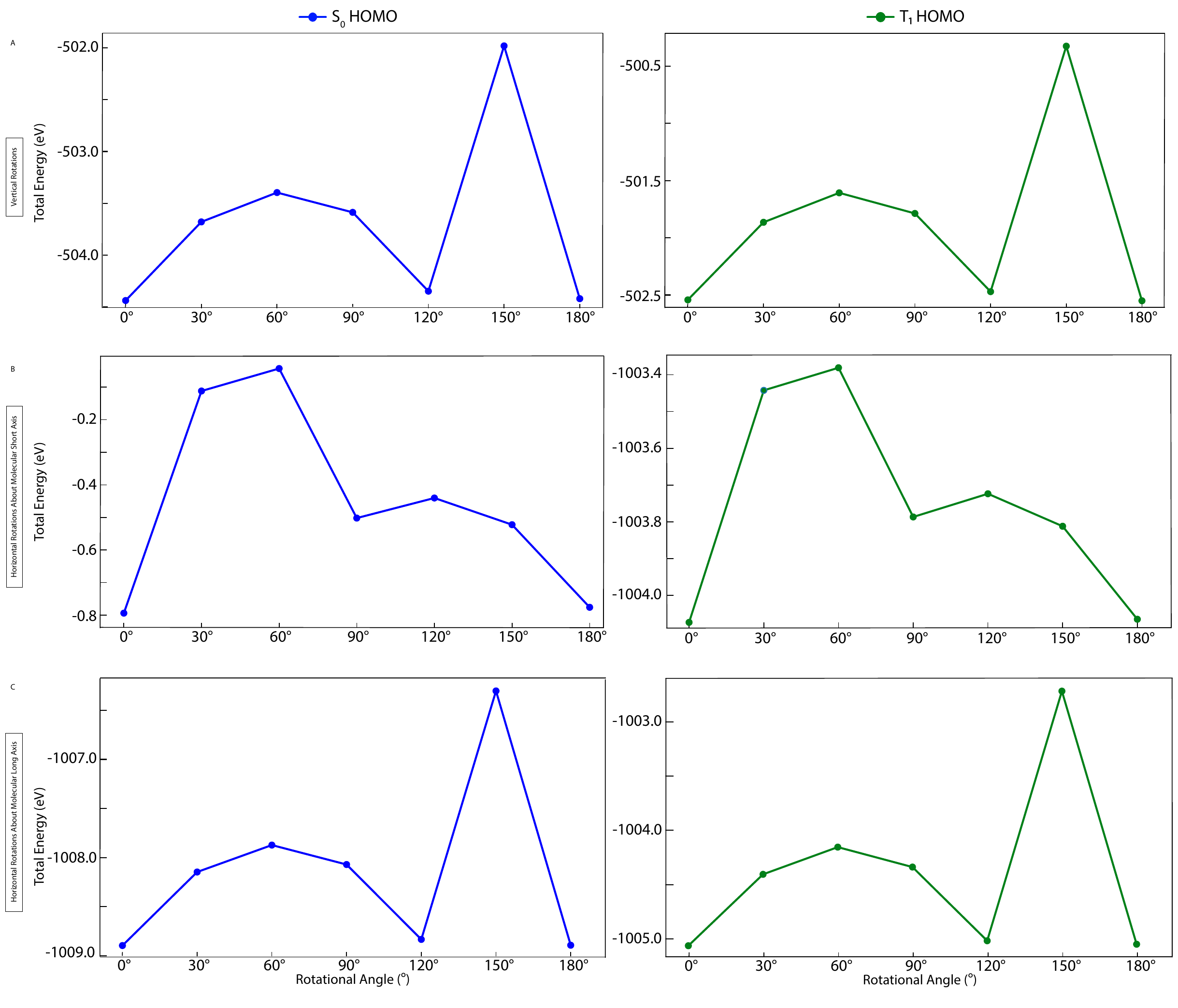}
    \caption{\T{Angular Dependence of Total Cocrystal Energy.} One molecule in the pentacene dimer is rotated clockwise along the (A) vertical, (B) horizontal nearest-neighbor along the in-plane molecular short axis (6.13 \AA~apart), and (C) horizontal nearest-neighbor along the in-plane, molecular long axis (4.10 \AA~apart). DHP molecules are omitted. The total energy of the structure is minimized at the intrinsic inter-pentacene configurations.}
    \zfl{SI_DFT_EO}
\end{figure*}

\begin{figure*}[t]
    \centering
    \includegraphics[width=0.72\textwidth]{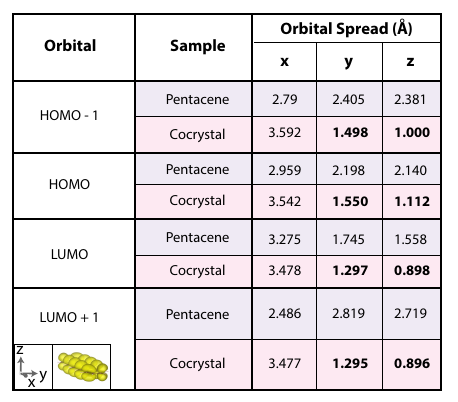}
\caption{\T{HOMO and LUMO orbital spread along pentacene molecular axes in the cocrystal and pure pentacene} reveal greater orbital localization along the y and z molecular axes (bolded) and greater delocalization along the molecular x axis, defined in the inset, for pentacene in DHP/Pc vs. a pure pentacene crystal.}
    \zfl{DFT_Spread}
    \label{DFT_Spread}
\end{figure*}

\end{document}